\documentclass[sigconf, screen]{acmart}

\AtBeginDocument{%
  }

\setcopyright{none} 
\copyrightyear{2026}
\acmYear{2026}
\acmDOI{XXXXXXX.XXXXXXX}

\acmISBN{978-X-XXXX-XXXX-X/XX/XX}

\usepackage{multirow}
\usepackage{graphicx}
\usepackage{lscape}
\usepackage{algorithm}
\usepackage{algpseudocode}
\usepackage{amsmath}

\usepackage{amssymb}
\usepackage{enumitem}
\usepackage{fixmath}
\usepackage{bm}
\graphicspath{{figures/}}
\usepackage{subcaption}
\usepackage{float}

\begin{document}

\title{Salca: A {\underline{S}parsity-\underline{A}ware} Hardware Accelerator for Efficient {\underline{L}ong-\underline{C}ontext \underline{A}ttention} Decoding}

\author{Wang Fan}
\affiliation{\institution{Fudan University} \city{Shanghai} \country{China}}

\author{Wei Cao}
\authornote{Corresponding Author} 
\affiliation{\institution{Fudan University} \city{Shanghai} \country{China}}
\email{caow@fudan.edu.cn}

\author{Xi Zha}
\affiliation{\institution{Fudan University} \city{Shanghai} \country{China}}

\author{Kedi Ma}
\affiliation{\institution{Fudan University} \city{Shanghai} \country{China}}

\author{Mingqian Sun}
\affiliation{\institution{Fudan University} \city{Shanghai} \country{China}}

\author{Jialin Chen}
\affiliation{\institution{Fudan University} \city{Shanghai} \country{China}}

\author{Fengzhe Zhang}
\affiliation{\institution{Fudan University} \city{Shanghai} \country{China}}

\author{Fan Zhang}
\authornote{Corresponding Author} 
\affiliation{\institution{Fudan University} \city{Shanghai} \country{China}}
\email{ffzhang@fudan.edu.cn}





\begin{abstract}
Long contexts improve capabilities of large language models but pose serious hardware challenges: compute and memory footprints grow linearly with sequence length. Particularly, the decoding phase continuously accesses massive KV cache, dramatically increasing bandwidth and computing pressure. Existing accelerators are primarily designed and evaluated for short contexts. They suffer from significant performance degradation when processing long contexts. To bridge this gap, we identify the major bottleneck and present a hardware accelerator for long context attention decoding via hardware-software co-design. On the software side, we propose dual-compression dynamic sparse attention. It combines ultra-low-precision quantization with feature sparsity to minimize prediction overhead. A hardware-friendly approximate Top-K selection further reduces filter complexity from $O(n \log k)$ to $O(n)$. On the hardware side, we deeply optimize compute and memory access to tackle bottlenecks from intricate interplay between sparse attention and long contexts, and establish a performance model to derive the optimal co-design scheme. The resulting hardware adopts a fully pipelined parallel architecture and achieves $O(n)$ efficiency even for long sequences. Experiments show that our design delivers $3.82\times$ speedup and $74.19\times$ energy efficiency over A100. Compared to SOTA accelerators, this is the first ASIC accelerator that efficiently supports long context inference, with at least $3.5\times$ higher throughput and $2.08\times$ better energy efficiency.

\end{abstract}




\maketitle

\section{Introduction}
Recent advancements have propelled large language models (LLMs) beyond simple tasks toward complex reasoning over extensive input sequences\cite{Minference-1}. For instance, in multimodal and document-intensive applications, context windows frequently span tens of thousands of tokens\cite{Quest}\cite{R-kv} \cite{Sampleattention}\cite{Sparsemm}. This evolution creates a stringent demand for long context reasoning capabilities\cite{infinigen} \cite{Flexprefill}. 

\begin{figure}[t]
\centering
\begin{subfigure}[b]{0.45\columnwidth}
\centering
\includegraphics[width=\linewidth]{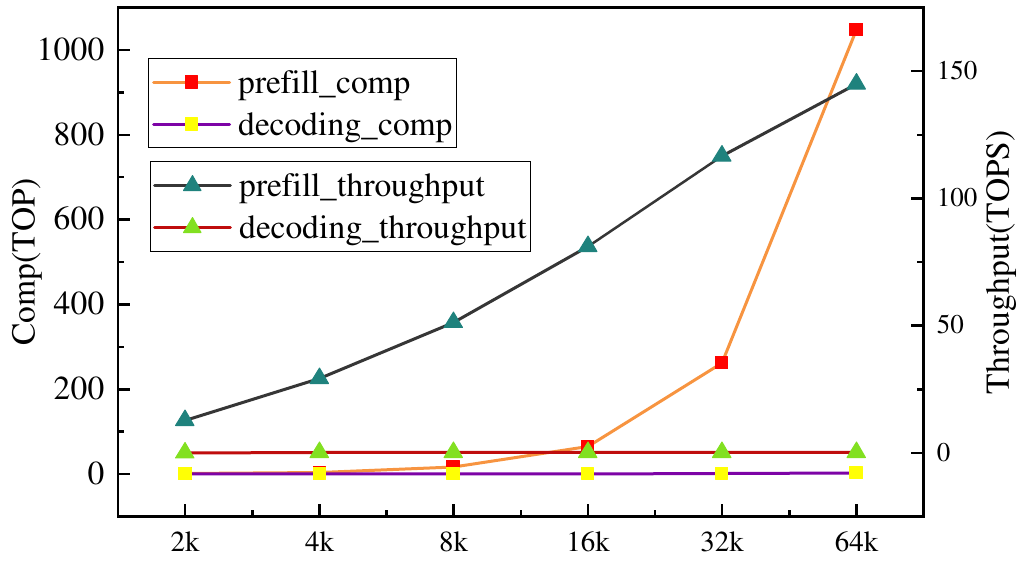}
\caption{Throughput Comparison}
\label{fig:Comp_analyze}
\end{subfigure}
\hspace{0.5em}
\begin{subfigure}[b]{0.40\columnwidth}
\centering
\includegraphics[width=\linewidth]{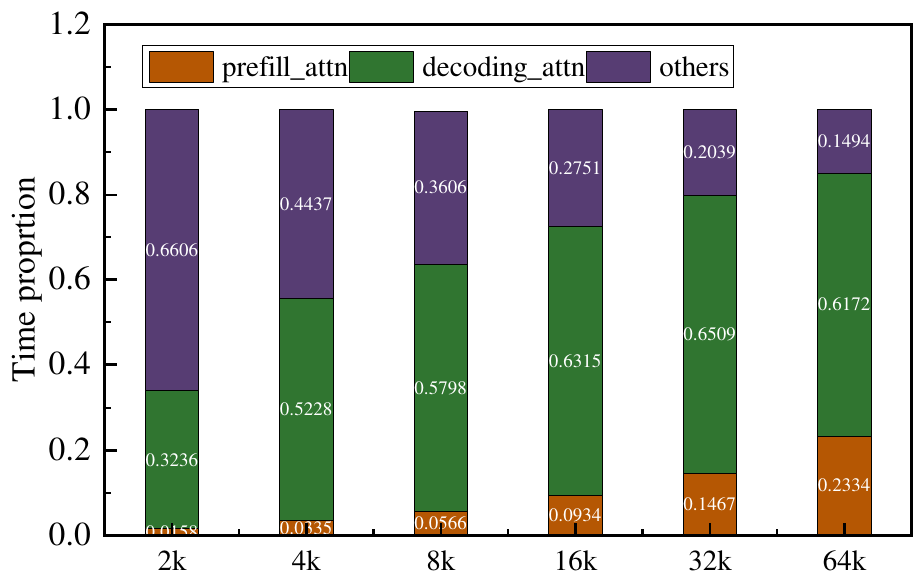}
\caption{Time analysis}
\label{fig:Time_propotion}
\end{subfigure}
\caption{Difference between prefilling and decoding.}
\label{fig:PD_compare}
\end{figure}

Transformer inference consists of prefilling and decoding, which differ fundamentally in computing patterns and bottlenecks\cite{Sarathi} \cite{splitwise} \cite{distserve}. Prefilling executes matrix multiplication, making it fully exploit GPU parallel capability\cite{Trinity} \cite{kamath2025pod}, as shown in Fig.\ref{fig:Comp_analyze}. In contrast, decoding is autoregressive generation phase. It only generates one token at each step but strictly requires a full scan of KV cache. Throughput is entirely limited by bandwidth\cite{spad} \cite{Ellie} \cite{Enhancing_PD} \cite{li2024llm}. In Fig.\ref{fig:Time_propotion}, we profile attention execution time when decoding 128 tokens. It reveals two insights. First, attention latency scales rapidly with sequence length, ultimately dominating over 80\% of total time. Second, decoding emerges as definitive performance bottleneck. For instance, at a 32K input length, decoding phase accounts for less than 1\% of total FLOPs yet consumes over 65\% of execution latency. Therefore, accelerating decoding in long context scenarios (LCS) remains central challenge for Transformer inference optimization.

Sparse attention provides a highly viable solution for efficient inference in LCS\cite{vasylenko2025long} \cite{Lserve} \cite{seerattention}. Attention distribution is inherently non-uniform. For a given query vector, its semantic information is only strongly correlated with a subset of K/V matrix\cite{tay2020sparse} \cite{Spargeattn}. Fig.\ref{fig:chatglm_sparsity_heatmap} profiles ChatGLM3 model on long-text tasks. When 95\% of attention probability is cumulatively covered, the average sparsity ratio reaches 94.6\%, with 70\% attention heads exceeding 95\%. This means that over 90\% of attention weights are redundant. This provides opportunities for aggressive computing compression. By exclusively focusing on critical subset, attention complexity drops from $O(n^2)$ to a near-linear level without sacrificing accuracy\cite{thangarasa2023sparse} \cite{zhang2025vision} \cite{liu2021transformer}. This property is particularly important for LCS. It breaks the constraint of loading all K/V during each decoding step, directly mitigating severe compute and bandwidth bottlenecks imposed by growing sequence lengths\cite{rao2021dynamicvit}.

Recently, various hardware-software co-designed accelerators have emerged to optimize sparse attention\cite{DOTA}\cite{DTATrans}. For example, SpAtten removes secondary information through cascaded pruning\cite{Spatten}; Energon introduces multi-round filtering and dynamic threshold\cite{Energon}; ELSA proposes hash mapping approximation\cite{ELSA}. However, these works are mainly designed and evaluated for short context scenarios (SCS), typically no longer than 4K. Transitioning to long context decoding will violate their underlying assumptions, exposing three critical challenges. First, escalating prediction overhead diminishes benefits of lightweight filtering. Taking 4-bit quantization schemes (adopted by Energon\cite{Energon} and Sanger\cite{Sanger}) as an example, data loading volume of filter stage is $2\times$ that of attention when sparsity is 95\%. This results in severe power costs. At a sequence length of 8K, filter stage already accounts for more than 60\% of total power, yielding only 25\% power savings compared with dense attention\cite{SOFA}. Second, sparse pattern selection introduces significant latency\cite{Spatten} \cite{DAC_1bit_quant} \cite{Energon}. Identifying sparsity relies on extracting Top-K elements from approximation score, which has $O(n \log k)$ complexity\cite{matsumoto2015optimal}. In SCS, $n$ and $k$ are small, so time overhead is not significant. However, in LCS, $n$ and $k$ increase substantially. Sorting performance deteriorates sharply, becoming critical path of end-to-end latency. Third, data supply becomes bottleneck. In SCS, KV cache can reside in on-chip SRAM. However, it far exceeds on-chip capacity during decoding in LCS. And, due to the lack of reuse, data must be frequently read from off-chip memory\cite{fu2024challenges}. This renders on-chip dataflow and reuse mechanisms ineffective and imposes higher bandwidth requirements.

To overcome these limitations, we propose Salca, an algorithm-hardware co-design scheme tailored for attention decoding in LCS. Anchored by sparse attention, Salca exclusively accesses the most relevant K/V vectors. Algorithmically, we introduce a dual compression method to minimize relevance estimation overhead. We exploit heavy-channel effect in LLMs to extract dominant features, aggressively pruning non-essential channels. Meanwhile, we push beyond the lower bound of traditional 4-bit quantization. Different quantization schemes are applied to Query and Key respectively to identify the minimum bit width that maintains ranking fidelity. The combined feature-precision compression reduces memory traffic to $\frac{1}{8}$ of 4-bit baselines. Furthermore, we optimize sparse pattern selection by eliminating costly search for exact $\operatorname{K}^{th}$ largest element. Instead, we utilize parallel statistical methods to approximate threshold, enabling Top-K filtering in $O(n)$ time complexity. 

On the hardware side, we propose a hardware accelerator with native support for decoding in LCS. Our design deeply co-optimizes compute and memory architecture. Computationally, we construct a five-stage pipeline, which has a linear $O(n)$ timing overhead even for extended sequences. An SRAM-based Top-K locating unit and a mechanism for sparse index extraction and dense store are designed specifically to resolve pipeline latency bottlenecks and speed mismatches. At memory level, we optimize HBM data layouts and introduce a reordering-based conflict resolution scheme to maximize transfer efficiency. Data is fetched directly from HBM and loading time is hidden through coordination of pipeline and memory accesses. To balance computing and memory, we further formulate a performance model to derive optimal computing-memory co-design scheme and corresponding resource allocation. 

The contributions of this paper can be summarized as follows:
\begin{itemize}[noitemsep, topsep=0pt, partopsep=0pt, parsep=0pt]
\item \textbf{An efficient sparse-pattern prediction mechanism.} We propose a novel relevance estimation scheme through dual compression to minimize memory access. Furthermore, we introduce an approximate Top-K mechanism with $O(n)$ complexity to reduce filter overhead.

\item \textbf{A hardware architecture for decoding in LCS.} We propose a fully pipelined architecture with highly optimized Top-K sorting and sparse mask store to eliminate hardware bottleneck. It also deeply optimizes data layout and memory access strategy for sparse access patterns. Additionally, a performance model is developed to balance the computing-memory and determine resource allocation.

\item \textbf{Performance Evaluation.} We evaluate Salca on 13 benchmarks. Compared to A100 GPU, Salca achieves 3.82$\times$ speedup and 74.19$\times$ energy efficiency gain. To best of our knowledge, this is the first ASIC accelerator that supports efficient long-context inference. In comparison with prior accelerators, it delivers at least 3.5$\times$ throughput improvement, along with gains of 1.33$\times$, 2.08$\times$, and 1.97$\times$ in core energy, device energy, and area efficiency, respectively.
\end{itemize}

\section{Background and Motivation}
\subsection{Quantization}
Quantization effectively reduces memory footprint and data-transfer pressure\cite{8-bit-quant} \cite{Awq} \cite{Smoothquant}. Quantization methods can be divided into asymmetric quantization and symmetric quantization\cite{Owq}. Asymmetric quantization fully exploits dynamic range to yield higher accuracy, albeit incurring complex quantization and dequantization overhead. Symmetric quantization fixes zero point, which limits representational capacity but significantly reduces computation.

\subsection{Sparse Attention}
\begin{figure}[!htbp]
\centering
\begin{subfigure}[b]{0.45\columnwidth}
\centering
\includegraphics[width=\linewidth]{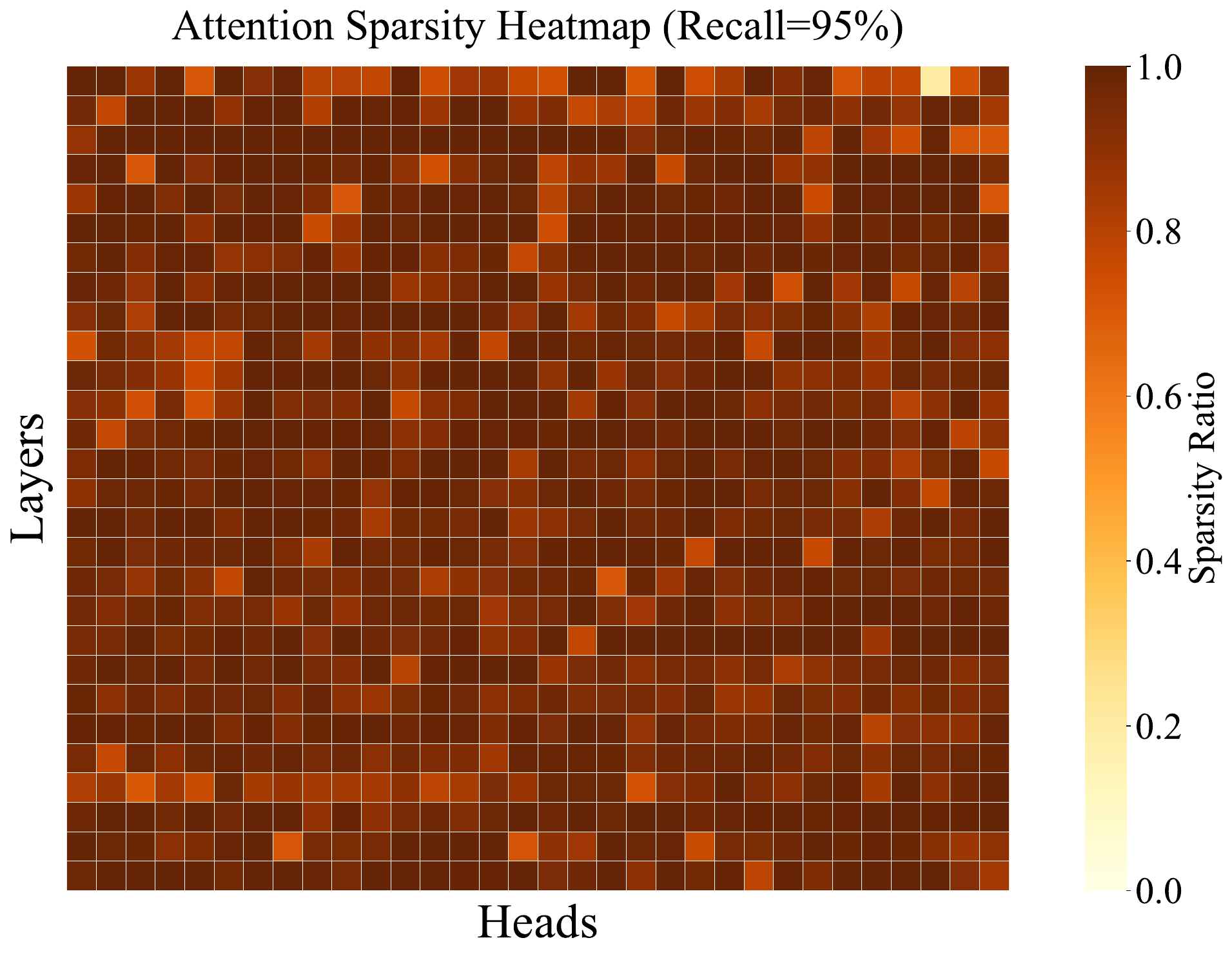}
\caption{Sparsity ratio in Chatglm3}
\label{fig:chatglm_sparsity_heatmap}
\end{subfigure}
\hfill
\begin{subfigure}[b]{0.48\columnwidth}
\centering
\includegraphics[width=\linewidth]{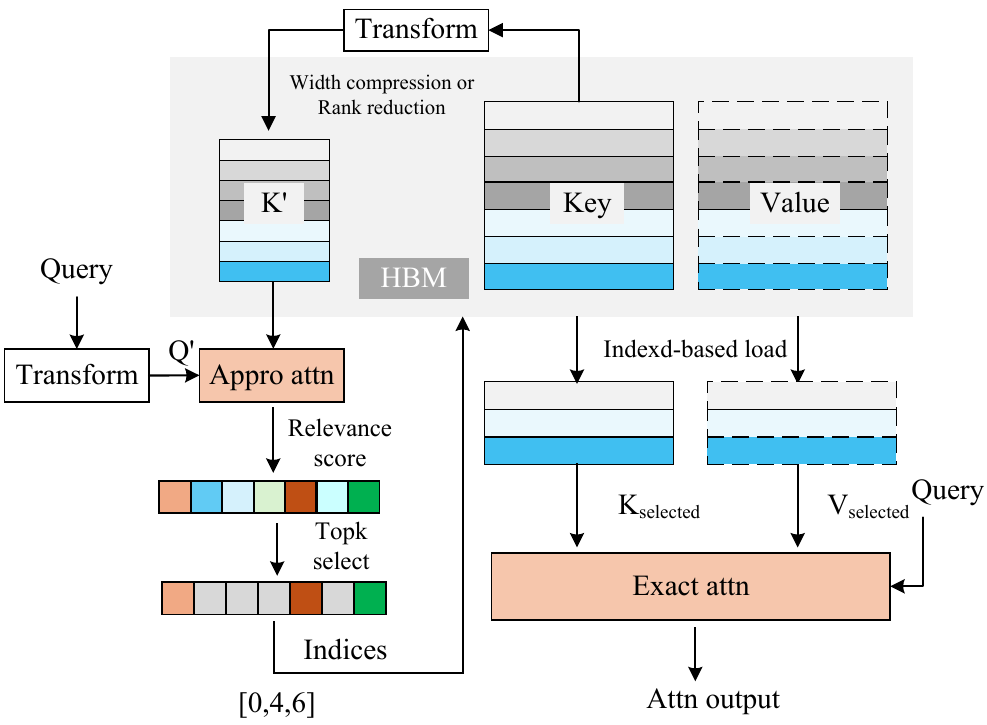}
\caption{Sparse attention framework}
\label{fig:Sparse_attention}
\end{subfigure}
\caption{Sparse attention.}
\label{fig:main}
\end{figure}

Practical sparse attention methods filter the most relevant tokens via low-overhead pre-computing process\cite{PADE} \cite{Pqcache}. As illustrated in Fig.\ref{fig:Sparse_attention}, it consists of three stages. First, estimate relevance between query and keys through approximate computing (e.g., low precision or low rank). Second, select sparse pattern based on relevance scores. The most common method is to directly select Top-K largest values and record their position indices. Third, execute exact attention computing only based on selected K/V vectors.

\subsection{Motivation}
Although sparse attention reduces complexity of decoding in LCS, theoretical gains are difficult to translate into practical acceleration. For GPUs, their execution model mismatches sparse computing patterns. For instance, SparQ yields mere 33\% attention speedup on A100\cite{Sparq}. Loki delivers only 40\% speedup\cite{Loki}. This performance gap stems from irregular memory accesses, degrading data fetching efficiency. Another problem of GPUs is limited support for ultra-low-precision or mixed-precision, such as 2/3-bit. This constrains optimization for relevance estimation in bit-width compression.

Hardware-software co-design provides an effective paradigm. Customizing computing and memory can resolve bottlenecks caused by irregularity of sparse attention. For example, SOFA\cite{SOFA} delivers $9.5\times$ speedup and $71.5\times$ better energy efficiency over A100. However, existing accelerators mainly target SCS, as detailed in Tab.\ref{tab:Comp_result_with_existing}. Adopting them for decoding in LCS presents multiple challenges.

\textbf{First, pre-computing becomes a significant computational burden.} Pre-computing overhead becomes pronounced due to two factors. First, models exhibit higher sparsity. Attention computing cost reduces relatively, while pre-computing still need to traverse entire context. Second, quantization enable attention to run at low precision (e.g., INT8)\cite{Atom}. Traditional pre-computing algorithm lose advantage. For instance, in Sanger, pre-computing uses 4-bit quantization. When attention precision drops from 16-bit to 8-bit, pre-computing power share surges from 37\% to 67\%, while overall power savings over dense computing fall from 61\% to 25\%\cite{SOFA}. Computational burden has shifted from attention to pre-computing. Therefore, a more lightweight pre-computing method is necessary. Crucially, relevance estimation merely requires accurate relative ranking, not absolute numerical precision. Based on this insight, we push quantization to aggressive sub-4-bit with feature sparsity to explore minimal compression scheme. This method slashes memory and computing cost simultaneously. This corresponds to Sec.\ref{sec:dual_compression}.

\textbf{Second, Top-K selection becomes critical bottleneck.} This stage involves serial comparisons with $O(n \log k)$ complexity. In LCS, increase of $n$ and $k$ degrades sorting performance sharply. Conversely, relevance estimation and attention have no data dependencies and can be processed in parallel with $O(n)$ complexity. Since pipeline is dictated by slowest stage, minimum processing latency remains bounded by $O(n \log k)$. This undermines theoretical benefits. We observe that dominant overhead stems from locating exact $\operatorname{K}^{th}$ largest element. However, sparse attention is fundamentally an approximate mechanism. Enforcing a strict Top-K threshold contradicts core philosophy. This insight motivates an approximate Top-K threshold locating method. We replace serial element sorting with independent categorization operations. Threshold Locating is transformed into statistical results analysis across a small set of categories. Complexity drops to $O(n)$. This corresponds to Sec.\ref{sec:topk_alg}.

\textbf{Third, data supply constraints degrade hardware utilization.} In LCS, KV cache memory footprint far exceeds on-chip SRAM, necessitating external memory such as HBM. Decoding process lacks data reuse, rendering cache strategies ineffective. Computing units must rely entirely on real-time streaming from external memory. Meanwhile, sparse attention degrades actual memory access efficiency in two aspects. First, discrete accesses and short burst transfers destroy spatial locality for memory reads. Second, parallel index-based access causes physical conflicts across memory channels. The two factors reduce HBM transfer efficiency from 95\% to 30\%. This throttles data supply and leaves computing units starving. Throughput also drops significantly. To tackle memory-bound nature, we argue for co-optimizing computing and memory paths. First, we should improve HBM transfer efficiency through optimized data layout and access strategies. Second, actual bandwidth, rather than theoretical bandwidth, should be evaluated as design constraint of computing units. Third, determine the maximal computing capability based on effective data supply capacity. This guarantees that hardware can get enough data while maximizing HBM bandwidth utilization. This corresponds to Sec.\ref{sec:hardware_architecture}.

\section{Algorithm Optimization}
\subsection{Dual Compression Optimization}\label{sec:dual_compression}
\noindent\textbf{Heavy-channel-based feature sparsity:} 
Feature dimensions of LLM attention heads exhibit significant redundancy. Prior work such as SparQ\cite{Sparq} and Loki\cite{Loki}  has demonstrated that retaining only a subset of features suffices to estimate relevance. However, existing schemes suffer from distinct limitations. SparQ relies on per-query feature selection, making feature positions vary across inputs. This destroys spatial locality for contiguous feature fetching. Loki achieves contiguous access through offline static extraction based on a calibration dataset, yet its generalization remains limited and performance degrades under high sparsity, as shown in Sec.\ref{sec:alg_eva}.

\begin{figure}
\centering
\includegraphics[width=0.7\columnwidth]{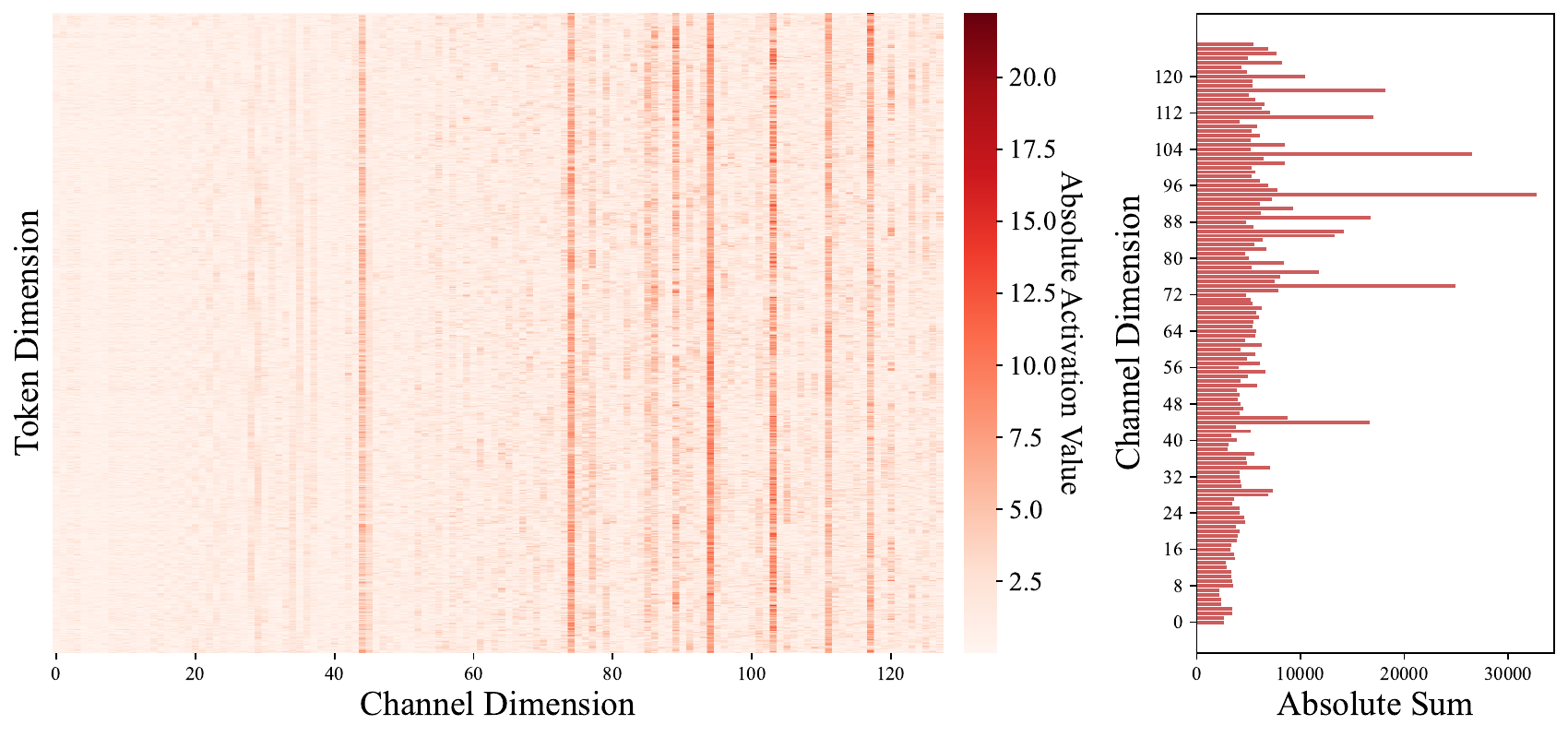}
\caption{Heavy channels of Key.}
\label{fig:heavy_channels}
\end{figure}

In LLMs, Key exhibits pronounced channel-wise characteristics: values vary little within channel but greatly across channels\cite{KIVI} \cite{Kvquant} \cite{Sageattention}. Based on this phenomenon, we propose an input-adaptive feature extraction. We identify channels with notably large magnitudes in Key, termed heavy channels. Due to larger magnitudes, heavy channels dominate dot product result. Relying solely on these channels suffices to measure token relevance. Moreover, heavy channels remain stable throughout inference, as shown in Fig.\ref{fig:heavy_channels}. It requires only a single identification for each input, and we can store heavy channel data as core features separately for continuous data loading. Specifically, let $s_f$ denote feature sparsity. We select $r = s_f \times d$ channels as heavy channels, where $d$ is head features. During prefilling, heavy channels are identified by reducing Key along token dimension: $S = \sum_{i=0}^{n-1} |\text{key}{[i,j]}|$. We then construct heavy channel index set by extracting top-$r$ channels: $\operatorname{I}_{heavy} = \operatorname{argTopk}(S, r)$. During decoding, relevance between query $q$ and key $k$ is estimated exclusively on this set: $\widetilde{A} = \sum_{j \in \operatorname{I}_{heavy}} q_j k_j$. 

\noindent\textbf{Minimum Quantization Bit-width Exploration:} 
To further mitigate computing costs, we explore minimum quantization width that preserves ranking fidelity. Given that Query and Key exert distinct impacts on computing and bandwidth, we abandon uniform quantization in favor of a two-stage strategy.

First, we minimize Key bit-width to alleviate memory pressure. Key is loaded from external memory, so its bit-width $w_k$ directly determines bandwidth consumption of pre-computing. We first fix Query at full precision and evaluate Key under different bit-widths and quantization strategies. Results in Tab.\ref{tab:mini_quant_width_test} show that $w_k = 1$ causes severe ranking distortion; although $w_k = 3$ brings slight improvement, it introduces $1.5\times$ memory access; when $w_k = 2$, asymmetric quantization already closely approaches full precision baseline. Therefore, we use 2-bit asymmetric quantization for Key.

Second, we optimize Query bit-width for computing efficiency. Query is stored on-chip, so increasing width $w_q$ does not affect memory access. We adopt symmetric quantization, as it incurs minimal dequantization overhead. Moreover, Query quantization factor is shared across all key vectors and can be omitted. Results show that $w_q = 3$ already meets sorting requirements, while $w_q = 4$ brings only marginal gains. Therefore, we use 3-bit symmetric quantization for Query. This breaks traditional 4-bit quantization barrier and reduces memory-access and computing overhead by $2\times$. When combined with feature sparsity, reduction can reach $8\times$.


\subsection{Efficient Sparse Pattern Selection}\label{sec:topk_alg}

Sparse pattern selection relies on extracting Top-K elements. Existing mechanisms face two challenges. First, conventional Top-K methods rely solely on value magnitudes, ignoring positional information. The selected K/V lack contextual coherence and therefore degrade model accuracy, as demonstrated in Li et al.\cite{Snapkv} and Yang et al.\cite{yang2024tidaldecode}. Second, exact Top-K sorting incurs $O(n \log k)$ complexity. This increases latency for long sequences and severely disrupts pipeline balance in hardware. To address these problems, we propose an accurate and efficient sparse pattern selection mechanism.

\noindent\textbf{Enhancing spatial locality:} 
Purely value-driven selection tends to ignore continuous semantic dependencies among tokens. Inspired by \cite{Snapkv}, we introduce a maxpooling with $stride=1$ when performing dynamic sparsification. This operation allows smaller values to be overwritten by larger neighbors. Consequently, positions surrounding high-score elements are co-selected, therefore enhancing locality. This operation can be bypassed when models already exhibit strong Top-K performance. Unlike in Li et al.\cite{Snapkv}, only maxpooling can work. Applying average pooling without softmax would dilute salient values with surrounding ones, thereby undermining selection accuracy.

\noindent\textbf{Approximate histogram-based Top-K filtering:} 
To improve sorting efficiency, we introduce an approximate Top-K selection method. We argue that sparse attention is inherently an approximate computing framework, which tolerates ranking errors. Exact $\operatorname{K}^{th}$ largest value is unnecessary. Prior methods locating precise thresholds incur substantial computing overhead. We instead propose an histogram filter mechanism. It includes three stages.

\textit{(1) Histogram generation:} We quantize FP16 scores to INT8, which can be regarded as a memory address. We then initialize an array of length $256$ and perform hit counting for quantized data.

\textit{(2) Threshold locating:} We accumulate array counts from high to low addresses. The first address $T$ whose cumulative count exceeds $K$ becomes approximate threshold.

\textit{(3) Parallel filtering:} We rescan quantized data and retain all elements $\geq T$.

This method completely eliminates comparison dependencies by independent categorization. It requires only two traversals. On a $64$-way parallel architecture, it can be done in $2n/64$ cycles.

Quantization preserves relative ordering, so elements belonging to Top-K in FP16 still remain in quantized Top-K. Different floating values may map to same integer during quantization, which expands Top-K set slightly. This issue only affects data equal to threshold, without affecting larger values. For $8$-bit quantization, each address contains $0.39\%$ of data on average. Therefore, increase in Top-K set size is about $0.19\%$, which is far smaller than common retention rate (e.g., $5\%$). Thus, the introduced overhead is negligible. To maximize computing efficiency, we place quantization before maxpooling and apply aforementioned sorting to pooled results. This enables replacing costly FP16 comparators in maxpooling with INT8 units. The overall process is shown in Algorithm\ref{alg:approx_attention}.

\renewcommand{\algorithmicrequire}{\textbf{Input:}}
\renewcommand{\algorithmicensure}{\textbf{Output:}}

\begin{algorithm}
\caption{Hardware-Efficient Approximate Attention Decoding}
\label{alg:approx_attention}
\begin{algorithmic}[1]
\Require High-precision Query vector $q \in \mathbb{R}^d$, Key/Value matrices $K, V \in \mathbb{R}^{N \times d}$, Heavy channels $\mathcal{I}_{\text{core}}$ (size $r$), 2-bit Key core features $K_r^{\text{2bit}} \in \mathbb{R}^{N \times r}$, Target sparse token count $k$
\Ensure Final attention output $o$

\Statex \textbf{// Phase 1: Lightweight Relevance Estimation}
\State $q_r^{\text{3bit}} \gets \text{Quant}(q[\mathcal{I}_{\text{core}}])$ \Comment{Extract features and quant}
\State $S \gets \text{Dequant} (q_r^{\text{3bit}} \times (K_r^{\text{2bit}})^T)$ \Comment{dot-product scores}

\Statex \textbf{// Phase 2: Quant for Topk Selection}
\State $S_{\text{INT8}} \gets \text{Quant}(S)$ \Comment{Hoist quantization before pooling}
\State $S_{\text{pool}} \gets \text{MaxPooling1D}(S_{\text{INT8}}, \text{stride}=1)$ \Comment{Maxpooling}

\Statex \textbf{// Phase 3: Approximate Histogram Filtering}
\State $C[0 \ldots 255] \gets 0$ \Comment{Initialize histogram array}
\For{$x$ \textbf{in} $S_{\text{pool}}$}
    \State $C[x] \mathrel{+}= 1$ \Comment{Histogram generation}
\EndFor

\State $acc \gets 0$, $T \gets 255$
\While{$acc < k$ \textbf{and} $T \geq 0$}
    \State $acc \mathrel{+}= C[T]$ \Comment{Reverse prefix sum to locate threshold}
    \State $T \gets T - 1$
\EndWhile

\State $T_{\text{top}} \gets T + 1$ \Comment{Global Top-K approximate threshold}
\State $\mathcal{I}_{\text{topk}} \gets \{i \mid S_{\text{pool}}[i] \geq T_{\text{top}}\}$ \Comment{Filter to obtain indices}

\Statex \textbf{// Phase 4: Exact Attention Computation}
\State $K_{\text{exact}} \gets \text{Gather}(K, \mathcal{I}_{\text{topk}})$, $V_{\text{exact}} \gets \text{Gather}(V, \mathcal{I}_{\text{topk}})$ \Comment{Fetch}
\State $P \gets \text{Softmax}\left(\dfrac{q \cdot K_{\text{exact}}^T}{\sqrt{d}}\right)$ \Comment{Compute attention probabilities}
\State $o \gets P \cdot V_{\text{exact}}$ \Comment{Compute final attention output}

\end{algorithmic}
\end{algorithm}

\section{Hardware Architecture}\label{sec:hardware_architecture}
\subsection{Architecture Overview}
\begin{figure}
\centering
\includegraphics[width=\columnwidth]{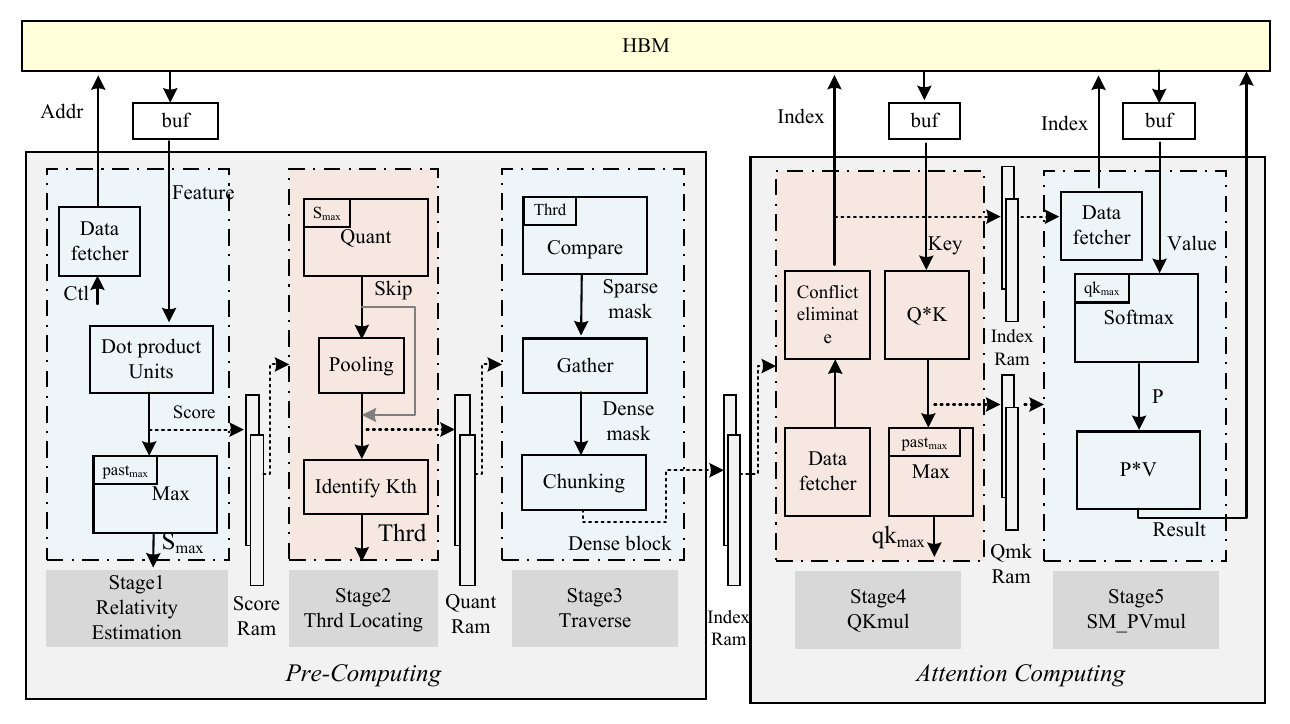}
\caption{Hardware Architecture Overview.}
\label{fig:Hardware_overview}
\end{figure}

To implement the algorithm, this paper proposes a novel architecture. As illustrated in Fig.\ref{fig:Hardware_overview}, it employs a five-stage pipeline. The first three stages generate sparse pattern indices, while the last two perform index-based attention. Data are loaded from HBM and hide latency through fine-grained pipeline. All stages operate in fully parallel. We remove data dependencies to achieve token-level parallelism within each stage. Meanwhile, we utilize double-buffer to isolate data access, enabling head-level parallelism between stages. This guarantees $O(n)$ linear complexity for decoding in LCS.

\noindent\textbf{Workflow:} The architecture activates once a new query is generated. Stage 1 computes relevance scores. It first issues HBM read requests to fetch core features. Low-precision dot-product units then calculate relevance scores. Results are stored in an on-chip Score RAM for subsequent stage. Stage 2 quantizes data in Score RAM to INT8. The quantized data then goes through MaxPooling unit or bypasses it. Results are streamed to Quant RAM and also fed to threshold locating unit, which can output $\operatorname{K}^{th}$ largest value in $n/p$ cycles, where $p$ is parallelism. Stage 3 performs filtering. It traverses Quant RAM and retains indices $\geq$ threshold. Stages 4 and 5 execute attention computing using generated indices. This process first eliminates channel conflicts during HBM access. It then computes attention and writes results back to HBM.

In following subsections, we will present our optimizations for computing and memory and introduce the performance model that balances the two.

\subsection{Computing Optimization}
Relevance estimation quantifies correlation between query and keys via dot product using a regular architecture of low-precision multipliers and adder trees. We omit it for brevity and focus on computing for Top-K selection and attention.

\subsubsection{\bfseries Multi-Level Reuse MaxPooling Unit}
Maxpooling extracts maximum values within sliding windows. Adjacent windows exhibit substantial overlap, particularly at $stride=1$. However, since overlap region changes dynamically, complete comparison results cannot be directly reused. To address this, we employ a multi-stage reuse strategy, which relies on the following recurrence:
\[
\begin{cases}
mp(r, n) = \max\big(mp(r-2, n-1), mp(r-2, n+1)\big) & r > 3 \\[2pt]
mp(3, n) = \max\big(in[n-1], in[n], in[n+1]\big) & r = 3
\end{cases}
\]
where $mp(r, n)$ is maxpooling result for window size $r$ ($r > 3$) centered at position $n$, spanning $[n - \lfloor r/2 \rfloor, n + \lfloor r/2 \rfloor]$, and $in[\cdot]$ is input. This formulation enables incremental computing of wide-range maxpooling through narrow-range intermediate results.

\begin{figure}[t]
\centering
\begin{subfigure}[b]{0.45\columnwidth}
\centering
\includegraphics[width=\linewidth]{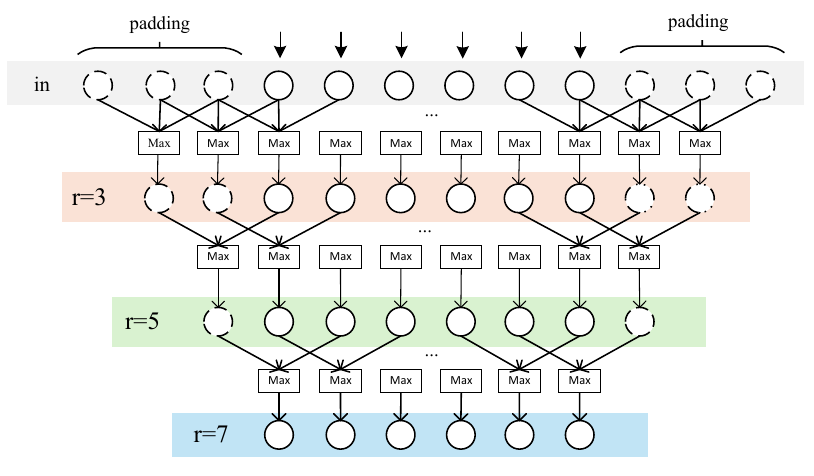}
\caption{Maxpooling Unit}
\label{fig:maxpoling_core}
\end{subfigure}
\hfill
\begin{subfigure}[b]{0.42\columnwidth}
\centering
\includegraphics[width=\linewidth]{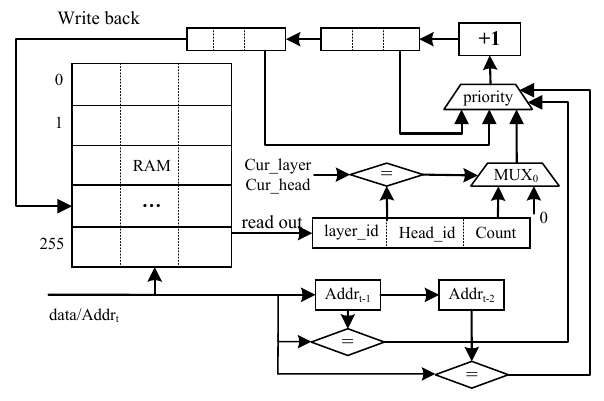}
\caption{Top-K Thrd Locating Unit}
\label{fig:topk_hardware}
\end{subfigure}
\caption{Computing Units in Threshold Locating Stage.}
\label{fig:main}
\end{figure}

Maxpooling unit is implemented as a hierarchical stride-comparison tree, as illustrated in Fig.\ref{fig:maxpoling_core}. We use $r = 7$ as an example. Input data is processed through three comparison stages. The first stage computes the maximum of every three consecutive elements, producing $mp(3, \cdot)$. Stages 2 and 3 each perform comparisons with a stride of 2, producing $mp(5, \cdot)$ and $mp(7, \cdot)$. By placing quantization prior to maxpooling, we replace FP16 comparators with lightweight INT8 units, significantly reducing hardware overhead.

\subsubsection{\bfseries SRAM-based Top-K Threshold Locating Unit}
To address the sorting requirement discussed in Section~\ref{sec:topk_alg}, we propose a SRAM-based Threshold Locating Unit. The unit centers on a pseudo-dual-port SRAM with auxiliary control logic, as shown in Fig.\ref{fig:topk_hardware}. Input data is directly transformed into SRAM read requests. The read value is accumulated and written back to same address, forming a read-accumulate-write pipeline. Different locating units have no data dependencies and can be fully parallelized. This effectively eliminates pipeline imbalance inherent in conventional Top-K selection. However, continuous input presents two challenges. First, SRAM contents must be flushed when processing different heads to prevent data pollution from historical counts. A global zeroing would incur a latency penalty of $256$ cycles. Second, histogram statistics rely on a read-write pipeline. Accessing same address consecutively triggers a Read-After-Write (RAW) hazard\cite{mahran2012handy}. To overcome these obstacles, we introduce two key mechanisms.

\noindent\textbf{Tag-based data isolation:} 
To achieve zero-overhead resetting, we employ a tag-based scheme to isolate stale data. Each SRAM entry is augmented with three fields $\langle \text{layer\_id}, \text{head\_id}, \text{count} \rangle$. Hardware compares tags of read entry against current pipeline's tag during processing. Any mismatch means read data is dirty, belonging to previous heads. In this case, a multiplexer discards read value and forwards a constant zero. Conversely, a full match confirms read data is valid for current head and will be forwarded. This dynamic ownership verification effectively prevents interference between distinct heads without explicit reset.

\noindent\textbf{Bypass-based hazard elimination:} 
RAW hazards occur when a subsequent request prematurely reads a location whose latest accumulated value has not yet been written back. To resolve this, we introduce a bypass mechanism with two delay registers, which record addresses and counts of the previous two requests. Concurrently with SRAM read, hardware compares current input address against addresses stored in delay registers. Any hit indicates the address was accessed recently. In this case, multiplexer selects count from delay registers as accumulator input. When both registers hit, the most recent one is selected with priority. This guarantees that every accumulation step operates on the newest value.

\subsubsection{\bfseries Traverse Unit}
\begin{figure}[]
\centering
\includegraphics[width=0.9\columnwidth]{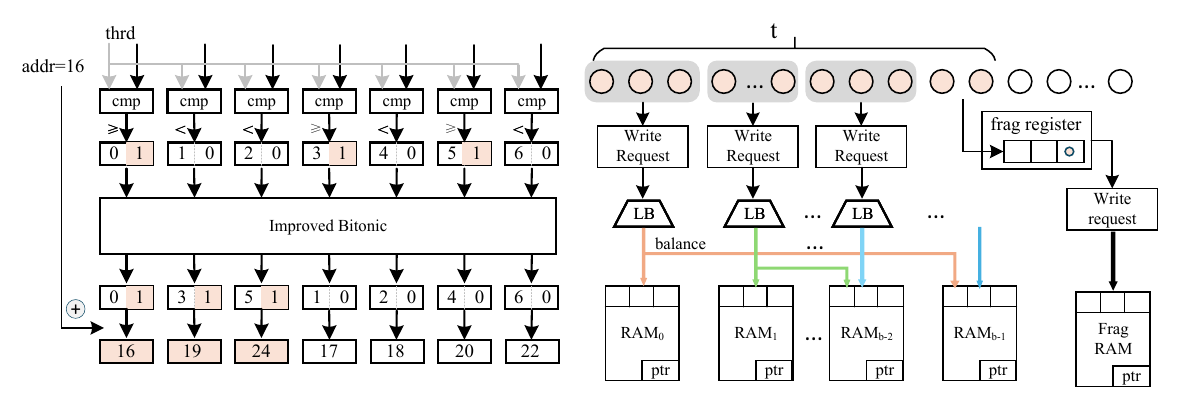} 
\caption{Index extraction and store.}
\label{fig:mask_store}
\end{figure}

Traverse stage filters valid elements ($\geq$ threshold) and generates indices. Index generation proceeds as a steady pipeline, while HBM reads based on indices stall frequently. This rate mismatch leads to an imbalanced producer-consumer\cite{thies2007practical}, necessitating an buffer to store indices. Implementing this buffer presents two challenges. First, valid elements are sparsely distributed with unpredictable positions, making extraction difficult. Second, valid element count varies per cycle. Data cannot be stored continuously like fixed length array. Storing data with gaps would inevitably introduce bubbles in subsequent memory access. To address these issues, we propose two mechanisms for efficient sparse index extraction and dense store, as shown in Fig.\ref{fig:mask_store}.

\noindent\textbf{Mask-sorting-based parallel gather:} 
We compare input elements against threshold to generate a binary mask, where $mask = 1$ denotes valid. Each element concatenates its local position with its mask and feeds into a modified Bitonic sorting network\cite{batcher1968sorting}. The network performs comparison solely on mask, while swapping position and mask together. This process compacts all valid elements to head of sequence. By adding base address to local position, absolute indices can be recovered. Unlike conventional zero eliminator, our design employs a fixed pipeline, yielding better hardware efficiency.

\noindent\textbf{Dense store for variable-length streams:} 
Since the number of valid elements $t$ per cycle constitutes a variable-length stream, we propose a chunk-based method for dense store. Index RAM is partitioned into $b$ banks, each capable of writing $s$ data per clock, satisfying $bs = p$, where $p$ is input number. Input data is divided into multiple chunks, with each corresponding to a bank. For each input, data is segregated into $\lfloor t/s \rfloor$ full chunks and a remainder of $t \% s$ fragments. Full chunks are written directly to associated banks, while fragments are temporarily held in a dedicated fragment register, which accumulates fragment. Once it collects $s$ elements, a write request to a specific Frag RAM is triggered. 

\subsubsection{\bfseries Attention Computing Unit}
\begin{figure}[]
\centering
\includegraphics[width=0.8\columnwidth]{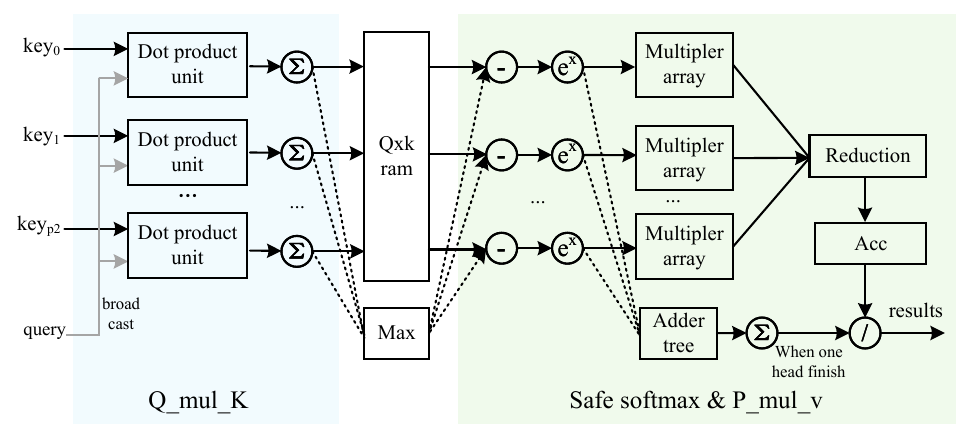} 
\caption{Exact attention computing unit.}
\label{fig:Attention_unit}
\end{figure}

Upon fetching data from HBM, attention computing is performed. As illustrated in Fig.\ref{fig:Attention_unit}, it is partitioned into two stages. The first stage computes segmented dot products and tracks maximum values. Complete dot product result $s_i$ requires accumulating partial sums over multiple cycles, as single pseudo-channel(PC) can only provide partial Key per cycle, as discussed in Sec.\ref{sec:data_layout}. Global maximum $\text{qk}_{\text{max}}$ is tracked for safe softmax. \cite{milakov2018online}. The second stage executes Online Softmax and Value multiplication using following formulation:
$\frac{\sum_{i=0}^{k} \left( e^{s_i - \text{qk}_{\text{max}}} \cdot V_i \right)}{\sum_{i=0}^{k} \left( e^{s_i - \text{qk}_{\text{max}}} \right)}$.

\subsection{Memory Optimization}
\subsubsection{\bfseries Data Layout}\label{sec:data_layout}
To address concurrent memory access demands, we optimize HBM layout. Benefiting from dual compression, storage overhead for pre-computing is significantly reduced. It is merely $\frac{1}{16}$ of K and V under $\frac{1}{4}$ feature sparsity. Given sequential read of pre-computing, we store core features contiguously. Conversely, K/V vectors require index-based access. We utilize the least significant bits to map to PC ID. Since HBM efficiency relies heavily on long bursts, we employ a single-PC mapping strategy rather than across-PCs for individual K/V distribution. Full data is obtained by combining results from multiple cycles. This maximizes burst length for better read efficiency.

\noindent\textbf{Interleaved access mechanism:} 
Pre-computing exhibits significant asymmetry: it demands high bandwidth but occupies minimal storage. This results in poor space utilization within allocated channels. In our architecture, utilizing one HBM2, pre-computing occupies $11$ PCs while attention occupies $16$. Although pre-computing saturates bandwidth of 11 PCs, it utilizes only $1/8$ of their capacity. This caps overall memory utilization at $70\%$, which is severe especially in LCS. We introduce an interleaved access mechanism to address this problem, as shown in Fig.\ref{fig:data_layout}. HBM is logically partitioned into two stacks, each serving pre-computing and attention. Each stack is further divided into two regions. Region0 consumes about $1/8$ capacity for core features, while Region1 consumes $7/8$ for K/V. It operates by alternating regions. Initially, system accesses core features in Stack0-Region0 and K/V in Stack1-Region1. Upon reaching capacity boundaries, configuration swaps, utilizing Stack0-Region1 and Stack1-Region0. This approach ensures that both HBM bandwidth and spatial utilization approach full saturation.

\subsubsection{\bfseries HBM Access Conflict Elimination}
\begin{figure}[t]
\centering
\begin{subfigure}[b]{0.50\columnwidth}
\centering
\includegraphics[width=\linewidth]{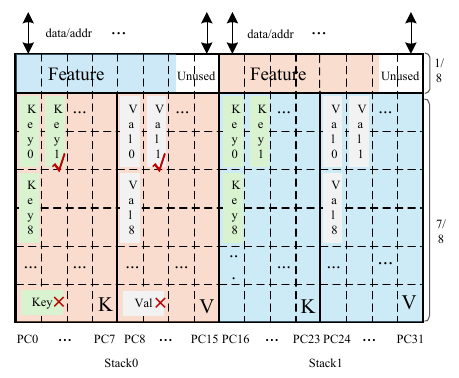}
\caption{HBM layout design.}
\label{fig:data_layout}
\end{subfigure}
\hfill
\begin{subfigure}[b]{0.8\columnwidth}
\centering
\includegraphics[width=\linewidth]{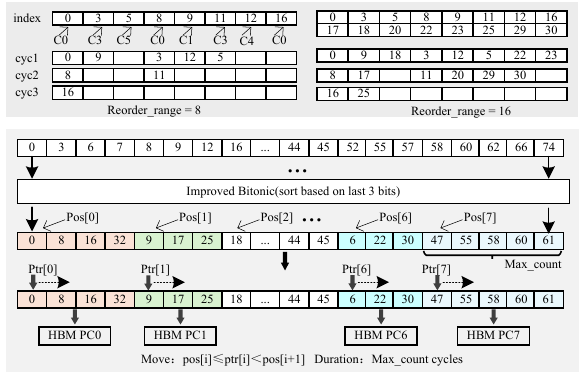}
\caption{Channel conflict elimination.}
\label{fig:channel_conflict}
\end{subfigure}
\caption{Memory Optimization.}
\label{fig:main}
\end{figure}

Top-K indices are used to load data from HBM. However, accessing HBM in order of index generation often triggers channel conflicts\cite{shi2022exploiting}, where multiple requests target same channel. A naive solution is to serialize conflicting requests into separate batches, but this results in significant channel under-utilization, as shown in Fig.\ref{fig:channel_conflict}. To address this, we propose a range-extension-based reordering scheme. By considering multiple accesses simultaneously, conflicts within a single request can be eliminated through merging and reordering with other requests. Consequently, only a minimal number of unresolvable conflicts require additional access.

Fig.\ref{fig:channel_conflict} illustrates the hardware architecture. We first map indices to corresponding channels. In our design, K/V data is distributed across $8$ PCs, so $3$ LSBs of index can identify target channel. We adopt improved Bitonic network to sort requests by PC ID. Accesses to same PC are clustered consecutively. We then record the maximum request count $\text{max\_count}$ across all channels, and starting position $\text{pos}[i]$ of each channel $i$ ($0 \le i \le 7$). Subsequently, each channel maintains an access pointer $\text{ptr}[i]$, initialized to its starting position $\text{ptr}[i] = \text{pos}[i]$. For following cycles, if $\text{pos}[i] \le \text{ptr}[i] < \text{pos}[i+1]$, the value pointed by $\text{ptr}[i]$ serves as valid access address for channel $i$ and pointer moves backward. Otherwise, access request for channel $i$ is invalid and pointer remains stationary. This process lasts for $\text{max\_count}$ cycles, during which all requests are scheduled without any conflicts. To evaluate reordering efficacy, we measured average conflict rate across various reordering ranges in Tab.\ref{tab:Channel_conflict_ratio}.  $128$ was chosen to balance hardware cost and elimination effect.

\begin{table}[t]
\centering
\setlength{\abovecaptionskip}{0pt}
\caption{Channel conflict ratio}
\label{tab:Channel_conflict_ratio}{%
\begin{tabular}{lclllll}
\hline
Reorder range & 8 & 16 & 32 & 64 & 128 & 256 \\ \hline
Conflict rate & 2.18 & 1.71 & 1.45 & 1.25 & 1.17 & 1.09 \\ \hline
\end{tabular}%
}
\end{table}

\subsection{Performance Model}\label{sec:performance_model}
Preceding optimizations target memory and computing independently, leaving balancing of the two processes unresolved. This balancing is reflected in allocation of HBM channels and parallelism of computing units. It also affects sparsity supported. For pre-computing, increasing channel and parallelism enhances filter capability, thereby enabling architecture to support higher sparsity. Conversely, if attention computing consumes more resources, filter capability of pre-computing is reduced, and the maximum sparsity decreases accordingly. To precisely characterize the relationship between channel count, parallelism, and sparsity, we propose a performance model. The model aims to find optimal compute-memory co-design scheme and resource allocation under given bandwidth and target sparsity constraints.

\begin{table}[t]
\centering
\setlength{\abovecaptionskip}{0pt}
\caption{Key Notations}
\label{tab:key_notations}
\resizebox{\columnwidth}{!}{
\begin{tabular}{cc}
\hline
Notation & Description \\ \hline
$\bm{d}$ & Head dimension \\
$\bm{chn}$ & HBM PC number \\
$\bm{bw}$ & HBM PC bandwidth \\
$\bm{n}$ & Sequence length \\
$\bm{s_q}$ & Sequence sparsity \\
$\bm{r_q}$ & Sequence retention rate, ${r_q} = 1 - {s_q}$\\
$\bm{s_f}$ & Feature sparsity \\
$\bm{\alpha}$ & Channel conflict ratio \\
$\bm{\beta_{pre}} / \bm{\beta_{att}}$ & HBM transfer efficiency of pre-computing/attention \\
$\bm{f_{\mathrm{CMP}}} / \bm{f_{\mathrm{HBM}}}$ & Clock frequency of computing / HBM \\
$\bm{h_{pre}} / \bm{h_{att}}$ & HBM PC number allocated to pre-computing/attention \\
$\bm{p_{pre}} / \bm{p_{att}}$ & Computing parallelism of pre-computing/attention \\
$\bm{m_{pre}} / \bm{m_{att}}$ & Memory access parallelism of pre-computing/attention \\
$\bm{u_{pre}} / \bm{u_{att}}$ & Hardware utilization of pre-computing/attention \\ \hline
\end{tabular}%
}
\end{table}

Decoding is memory-bound, and throughput hinges on I/O bandwidth. Therefore, our strategy is to first estimate the maximum data supply capability and then scale computing units under this constraint. We select memory access parallelism of pre-computing and attention, denoted as $m_{pre}$ and $m_{att}$, as important design parameters. They represent data quantity that pre-computing and attention intend to obtain respectively from HBM, thereby determining allocation of HBM PCs. For clarity, we define relevant variables in Tab.\ref{tab:key_notations}. $m_{pre}$ and $m_{att}$ are primarily constrained by HBM bandwidth. In pre-computing, accessing each key consumes $(2 \cdot d \cdot s_f + 32)$ bits, including multiple 2-bit features and two FP16 quantization factors. In attention, K and V are quantized to INT8, accessing each K and V pair requires approximately $2 \cdot (8 \cdot d)$. Total bandwidth demand per cycle is $B_{\mathrm{req}} = (2 \cdot d \cdot s_f + 32) \cdot m_{pre} + 16 \cdot d \cdot m_{att}$. Peak bandwidth provided by HBM is $B_{\mathrm{HBM}} = \mathrm{bw} \cdot chn$. To ensure that data demand does not exceed HBM supply capacity, bandwidth constraint is formulated as $[(2 \cdot d \cdot s_f + 32) \cdot m_{pre} + 16 \cdot d \cdot m_{att}] \cdot f_{\mathrm{CMP}} \le \mathrm{bw} \cdot c \cdot n \cdot f_{\mathrm{HBM}}$, where $f_{\mathrm{CMP}}$ and $f_{\mathrm{HBM}}$ are clock frequencies of computing units and HBM, respectively. They may work in different clock domains. Furthermore, we can infer the number of PCs allocated to each stage: $h_{pre} = \frac{(2 \cdot d \cdot s_f + 32) \cdot m_{pre} \cdot f_{\mathrm{CMP}}}{\mathrm{bw} \cdot f_{\mathrm{HBM}}}$ and $h_{att} = \frac{16 \cdot d \cdot m_{att} \cdot f_{\mathrm{CMP}}}{\mathrm{bw} \cdot f_{\mathrm{HBM}}}$, with constraint $h_{pre} + h_{att} \le \mathrm{chn}$.

$m_{pre}$ and $m_{att}$ also affect pipeline balance. Assume target sparsity is $s_q$, then target retention rate $r_q = 1 - s_q$. When pipeline is running stably, ideal processing time for pre-computing is $T_{pre}^{\text{ideal}} = \frac{n}{m_{pre}} + C_{pre} \approx \frac{n}{m_{pre}}$, where $\frac{n}{m_{pre}}$ is data loading time and $C_{pre}$ is pipeline depth, which is negligible compared to the former. In practice, even continuous HBM access does not achieve $100\%$ transmission efficiency. Actual latency is approximately $T_{pre}^{\text{actual}} \approx \frac{n}{\beta_{pre} \cdot m_{pre}}$. Similarly, ideal processing time for attention is $T_{att}^{\text{ideal}} \approx \frac{n \cdot r_q}{m_{att}} + C_{att} \approx \frac{n \cdot r_q}{m_{att}}$. However, actual hardware performance is affected not only by HBM transfer efficiency but also by channel conflict. Channel contention increases access latency by a multiplicative factor $\alpha$. Actual latency becomes $T_{att}^{\text{actual}} \approx \frac{n \cdot r_q \cdot \alpha}{\beta_{att} \cdot m_{att}}$. Unlike pre-computing, attention performs non-consecutive transfers, resulting in lower efficiency: $\beta_{att} < \beta_{pre}$. Overall pipeline latency is determined by critical path, denoted as $T = \max(T_{pre}^{\text{actual}}, T_{att}^{\text{actual}})$. After transformation, when $r_q < \frac{\beta_{att} \cdot m_{att}}{\beta_{pre} \cdot m_{pre} \cdot \alpha}$, $T_{pre}^{\text{actual}} > T_{att}^{\text{actual}}$, making pre-computing the bottleneck. When $r_q > \frac{\beta_{att} \cdot m_{att}}{\beta_{pre} \cdot m_{pre} \cdot \alpha}$, $T_{pre}^{\text{actual}} < T_{att}^{\text{actual}}$, making attention the bottleneck. Therefore, $r_q = \frac{\beta_{att} \cdot m_{att}}{\beta_{pre} \cdot m_{pre} \cdot \alpha}$ represents minimum retention rate supported by architecture, which must approach target retention rate to meet design requirements. Combining this with bandwidth constraints yields Pareto frontier for $m_{pre}$ and $m_{att}$, which represents maximum data supply capability.

We further investigate computing parallelism for pre-computing and attention. Let $m_{pre}'$ and $m_{att}'$ be a Pareto optimal solution. Ideally, $p_{pre} = m_{pre}'$ and $p_{att} = m_{att}'$, allowing data loading and computing to match perfectly. However, data transfer efficiency falls short of $100\%$, and particularly, $\beta_{att}$ is far lower. Delayed data leads to hardware idling, making $u_{pre} = \beta_{pre}$ and $u_{att} = \beta_{att}$. To improve hardware utilization, parallelism of computing units must match the amount of data actually fetched: $p_{pre}' = \lceil m_{pre}' \cdot \beta_{pre} \rceil$, $p_{att}' = \lceil m_{att}' \cdot \beta_{att} \rceil$, where $\lceil \cdot \rceil$ denotes ceiling function. Consequently, utilization improves to $u_{pre} = \frac{m_{pre}' \cdot \beta_{pre}}{\lceil m_{pre}' \cdot \beta_{pre} \rceil} > \beta_{pre}$ and $u_{att} = \frac{m_{att}' \cdot \beta_{att}}{\lceil m_{att}' \cdot \beta_{att} \rceil} > \beta_{att}$. The obtained $p_{pre}'$ and $p_{att}'$ may be irregular integers, which complicates hardware design such as Bitonic networks. We can fine-tune them under conditions $p_{pre} \le p_{pre}'$ and $p_{att} \le p_{att}'$ to ensure sufficient data supply. After determining $p_{pre}$ and $p_{att}$, $m_{pre}$, $m_{att}$, $h_{pre}$, and $h_{att}$ can be derived as $m_{pre} = \lceil p_{pre} / \beta_{pre} \rceil$, $m_{att} = \lceil p_{att} / \beta_{att} \rceil$, $h_{pre} = \big\lceil (2 d s_f + 32) \cdot p_{pre} \cdot f_{\mathrm{CMP}} / (\beta_{pre} \cdot \mathrm{bw} \cdot f_{\mathrm{HBM}}) \big\rceil$, $h_{att} = \big\lceil 16 d \cdot p_{att} \cdot f_{\mathrm{CMP}} / (\beta_{att} \cdot \mathrm{bw} \cdot f_{\mathrm{HBM}}) \big\rceil$, which dictate HBM bandwidth allocation.

The above process provides a method for designing hardware that match different sparsity, memory, and computing. In our design, we adopt one HBM2 with $32$ PCs. HBM runs at $1$ GHz, while computing units run at $500$ MHz. We aim to maximize $m_{pre}$ and $m_{att}$. Empirical measurements yield $\alpha \approx 1.17$, $\beta_{pre} \approx 95\%$, and $\beta_{att} \approx 55\%$. We set $s_f = \frac{1}{2}$, $s_q = 95\%$ and $r_q = 5\%$, which is sufficient for most long context scenarios. We obtain $m_{pre} = 25$, $m_{att} = 2$, $h_{pre} = h_{att} = 16$, $p_{pre} = 24$, $p_{att} = 2$, achieving a minimum retention rate of $3.9\%$. To simplify hardware design, we set $p_{pre} = 16$. Meanwhile, given that $m_{att} \cdot \beta_{att} = 1.1 \ll 2$, we set $p_{att} = 1$ with a buffer for higher hardware utilization. This configuration supports a minimum retention rate of $5.8\%$, allocating $11$ PCs to pre-computing and $16$ to attention.

\section{Evaluation}

\subsection{Algorithm Evaluation}\label{sec:alg_eva}
\textbf{Tasks:} LongBench\cite{Longbench} is a standard long context understanding benchmark spanning Single-Doc QA, Multi-Doc QA, Summarization, Few-shot Learning, Code Completion, and Synthetic Tasks. With an average sequence length of 10K and maximum approaching 40K, it is widely adopted for assessing long context reasoning performance\cite{Snapkv} \cite{Pqcache} \cite{Pyramidkv} \cite{double_sparsity} \cite{KIVI} \cite{Kvquant}.

\noindent\textbf{Models:} 
We select three models from official LongBench leaderboard: LongChat-v1.5-7B-32k\cite{Longchat}, Vicuna-v1.5-7B-16k\cite{Vicuna}, and ChatGLM3-6B-32k\cite{ChatGLM3}. These models support maximum sequence lengths of 32K, 16K, and 32K, respectively. Head dimension is 128.

\noindent\textbf{Baselines:} We compare our method against following baselines.
\begin{itemize}[noitemsep, topsep=0pt, partopsep=0pt, parsep=0pt]
\item \textbf{Std\_TopK:} standard Top-K which computes precise attention scores and selects Top-K elements.
\item \textbf{H2O:} static sparsity using accumulated attention scores for token retention\cite{H2o}.
\item \textbf{SnapKV:} static sparsity based on observation windows at sequence ends with pooling\cite{Snapkv}.
\item \textbf{Loki:} offline feature selection via calibration datasets for key rank reduction\cite{Loki}.
\item \textbf{MoBA:} generates block-level representation vectors and selects the most relevant blocks\cite{Moba}.
\end{itemize}

\noindent\textbf{Accuracy Evaluation: } 
To evaluate accuracy of the proposed method, we conducted comparative experiments against various baselines on LongBench. Attention is executed under 8-bit quantization. For SnapKV, we set observation window to $32$ and pooling range to $7$. For MoBA,  block size is $4$. For Loki and our method, we evaluate performance under feature sparsity of $1/4$, $3/8$, and $1/2$, and report the best accuracy. Tab.~\ref{tab:accuracy_comparison} summarizes accuracy results for retaining only top-$1024$ with average retention rate of $9.4\%$. We applied maxpooling for LongChat and Vicuna and omitted it for ChatGLM3, given its strong Top-K performance.

\begin{table*}[!htbp]
\centering
\setlength{\abovecaptionskip}{0pt}
\caption{Accuracy comparison across different models and methods.}
\resizebox{\textwidth}{!}{%
\begin{tabular}{llcccccccccccccl}
\hline
 &  & NrtvQA & Qasper & MF-en & HotpotQA & 2WikiMQA & Musique & GovReport & QMSum & VcSum & TREC & TriviaQA & PR-en & RB-p & avg \\ \hline
\multicolumn{1}{c}{\multirow{8}{*}{LongChat-v1.5-7B-32k}} & Full & \textbf{20.95} & \textbf{29.73} & \textbf{43.35} & \textbf{33.05} & \textbf{24.12} & \textbf{14.11} & \textbf{31.14} & \textbf{23.05} & \textbf{6.49} & \textbf{67} & \textbf{79.60} & \textbf{31.5} & \textbf{56.92} & 35.46 \\ \cline{2-16} 
\multicolumn{1}{c}{} & Std\_topk & 18.89 & 27.77 & 29.51 & 29.33 & 23.07 & 11.51 & 27.15 & 22.57 & 2.81 & 66.5 & 56.81 & 25.5 & 52.44 & 30.30 \\
\multicolumn{1}{c}{} & H2O & 15.33 & 23.48 & 26.82 & 29.13 & 23.54 & 7.73 & 14.48 & 17.81 & 0.93 & 57.5 & 55.89 & 16.45 & 38.95 & 25.23 \\
\multicolumn{1}{c}{} & Loki & 15.53 & 24.22 & 28.85 & 18.89 & 21.05 & 7.42 & 25.07 & 20.89 & \textbf{8.28} & 69 & 59.82 & 4.50 & 37.88 & 26.26 \\
\multicolumn{1}{c}{} & SnapKV & 17.93 & 26.3 & 38.99 & 35.47 & 23.91 & 13.38 & 23.27 & 22.35 & 5.47 & 64 & 74.14 & 30 & 56.72 & 33.23 \\
\multicolumn{1}{c}{} & Moba & 15.28 & 28.68 & 36.26 & 27.24 & 23.67 & 8.39 & 23.65 & 20.83 & 4.18 & 63.50 & 61.90 & 17.68 & 49.41 & 29.28 \\
\multicolumn{1}{c}{} & Pl\_topk & 20.32 & \textbf{29.73} & \textbf{40.1} & \textbf{36.74} & 23.50 & \textbf{13.98} & \textbf{30.22} & 22.66 & 6.56 & 66 & 75.88 & 30 & 56.77 & 34.80 \\
\multicolumn{1}{c}{} & Salca & \textbf{20.69} & 29.62 & 40.03 & 36.59 & \textbf{24.39} & 13.90 & 30.06 & \textbf{22.85} & 6.34 & \textbf{67} & \textbf{77.61} & \textbf{30.50} & \textbf{57.40} & \textbf{35.15} \\ \hline
\multirow{8}{*}{Vicuna-v1.5-7B-16k} & Full & 18.89 & 26.08 & 37.71 & 26.67 & 21.04 & 8.7 & 27.78 & 22.50 & 15.32 & 69.50 & 77.42 & 4.50 & 40.71 & 30.52 \\ \cline{2-16} 
 & Std\_topk & 15.94 & 23.39 & 26.4 & 18.79 & 21 & 7.71 & 23.29 & 20.42 & 5.95 & 67.5 & 57.83 & 4.50 & 36.41 & 25.32 \\
 & H2O & 5.26 & 19.40 & 18.53 & 18.14 & 19.4 & 2.70 & 10.52 & 10.14 & 4.28 & 36 & 34.85 & 4.50 & 28.21 & 16.30 \\
 & Loki & 15.53 & 24.22 & 28.85 & 18.89 & 21.05 & 7.42 & 25.07 & 20.89 & 8.28 & 69 & 59.73 & 4.50 & 37.88 & 26.25 \\
 & SnapKV & \textbf{19.33} & 25.28 & \textbf{35.64} & 23.33 & 21.44 & 6.47 & 21.78 & 22.49 & 12.18 & 66.5 & 67.12 & 4.50 & 39.16 & 28.09 \\
 & Moba & 14.85 & 23.93 & 28.45 & 22.65 & 19.79 & 6.49 & 26.61 & 22.46 & 12.65 & 68 & 63.23 & 4.75 & 36.86 & 26.98 \\
 & Pl\_topk & 18.53 & 26.39 & 35.03 & 24.01 & 20.99 & 7.86 & \textbf{28.02} & 22.63 & \textbf{13.63} & 69 & 66.07 & 4.50 & \textbf{40.21} & 28.99 \\
 & Salca & 18.47 & \textbf{26.83} & 35.49 & \textbf{24.43} & \textbf{21.95} & \textbf{8.26} & 27.61 & \textbf{23.24} & 13.6 & \textbf{69} & \textbf{72.20} & 5.50 & 39.88 & \textbf{29.73} \\ \hline
\multirow{7}{*}{ChatGLM3-6B-32k} & Full & 26.08 & 43.28 & 51.66 & 55 & 44.42 & 40.38 & 36.68 & 24.06 & 17.85 & 79 & 83.70 & 99 & 54.12 & 50.40 \\ \cline{2-16} 
 & Std\_topk & 26.12 & 43.71 & \textbf{51.44} & 54.23 & 46.57 & \textbf{38.98} & 36.73 & 24.35 & 17.18 & 79 & 83.70 & 99 & 53.59 & 50.35 \\
 & H2O & 25.49 & 32 & 40.18 & 51.68 & 41.28 & 31.94 & 29.84 & 22.54 & 14.70 & 68 & 85.11 & 97 & 53.41 & 45.63 \\
 & Loki & 25.73 & \textbf{43.74} & 50.97 & \textbf{57.66} & 44.08 & 38.07 & \textbf{36.86} & 24.26 & \textbf{17.66} & 79 & 84.23 & \textbf{99.5} & 53.04 & \textbf{50.37} \\
 & SnapKV & 27.48 & 39.15 & 47.72 & 53.48 & 45.67 & 38.57 & 30.63 & 23.45 & 14.40 & 75.50 & \textbf{87.73} & 99 & \textbf{53.73} & 48.96 \\
 & Moba & 26.59 & 42.76 & 50.23 & 54.14 & 46.48 & 38.03 & 35.86 & 24.18 & 17.20 & 79 & 83.52 & 99 & 53.68 & 50.05 \\
 & Salca & \textbf{27.68} & 42.72 & 51.32 & 55.23 & \textbf{46.65} & 39.84 & 36.09 & \textbf{24.53} & 16.27 & \textbf{79} & 83.53 & 98 & 53.63 & 50.35 \\ \hline
\end{tabular}%
}
\label{tab:accuracy_comparison}
\end{table*}

A comprehensive comparison indicates that our method achieves superior performance in accuracy. Based on Tab.\ref{tab:accuracy_comparison}, we draw two conclusions. \textcircled{1} \textbf{Maxpooling and dynamic sparsity enhance stability of Top-K selection.} Experiments show that relying solely on traditional Top-K (e.g., H2O, Std\_TopK, and Loki) results in significant accuracy degradation for LongChat and Vicuna. SnapKV recovers partial accuracy, yet it fails when critical information drifts from sequence ends (e.g., gov\_report). MoBA employs a coarse-grained approach, analogous to pooling with $stride>1$. While incorporating locality, increased stride compromises flexibility and precision of selection, resulting in reduced accuracy. In contrast, our method deeply integrates fine-grained dynamic sparsity with maxpooling, balancing global selection and spatial locality for substantial accuracy gains. \textcircled{2} \textbf{Feature sparsity and ultra-low-precision quantization are orthogonal and dual compression does not compromise selection accuracy.} To evaluate the impact of our compression strategy, we compared it against uncompressed baseline, which keeps full features and precision in pre-computing. The baseline corresponds to Std\_TopK for ChatGLM and Pl\_TopK for LongChat and Vicuna. Pl\_TopK integrates maxpooling into Std\_TopK to align with our method. Results show that even with deep compression, our method still achieves accuracy close to the uncompressed baseline. The method also exhibits robustness in high sparsity scenarios; for instance, in NarrativeQA, model remains stable with average retention rate of only $5.5\%$. This indirectly confirms the redundancy in conventional methods relying on full features and 4-bit quantization.

\noindent\textbf{Feature Selection Evaluation:} 
To evaluate our heavy channel-based input-adaptive feature selection, we compared it with representative offline scheme, Loki. For LongChat and Vicuna, we constructed an enhanced baseline, Pl\_Loki, by integrating maxpooling into Loki. For ChatGLM3, we compare directly with Loki. Tab.\ref{tab:loki_comparison} details accuracy across varying feature sparsity. Loki suffers severe accuracy degradation at high feature sparsity and exhibits high variance across datasets. This instability stems from inaccurate feature selection when actual inputs deviate from calibration data. Our method dynamically selects heavy channels and has stronger input awareness capability. The last two columns report two metrics for the two methods: overlap with  true Top-1024 and coverage to true Top-512. Our method achieves higher overlap and better coverage on important tokens. Thus, it maintains stable performance across diverse datasets and high sparsity scenarios.

\begin{table*}[!htbp]
\centering
\setlength{\abovecaptionskip}{0pt}
\caption{Feature selection accuracy comparison}
\label{tab:loki_comparison}
\resizebox{\textwidth}{!}{%
\begin{tabular}{cccccccccccccccccc}
\hline
1 &  & NrtvQA & Qasper & MF-en & HotpotQA & 2WikiMQA & Musique & GovReport & QMSum & VcSum & TREC & TriviaQA & PR-en & RB-p & avg & Overlap & Coverage \\ \hline
\multirow{6}{*}{LongChat-v1.5-7B-32k} & Pl\_loki\_1/4 & 12.82 & 28.64 & \textbf{40.36} & 30.51 & \textbf{24.59} & 10.28 & 24.03 & 20.96 & 2.94 & 64 & 70.2 & 16.5 & 44.65 & 30.04 & 0.57 & 0.62 \\
 & Salca\_1/4 & \textbf{20.17} & \textbf{29.62} & 39.99 & \textbf{35.43} & 24.19 & \textbf{13.31} & \textbf{28.97} & \textbf{22.38} & \textbf{5.66} & \textbf{66} & \textbf{77.04} & \textbf{28.5} & \textbf{56.79} & \textbf{34.47} & 0.62 & 0.68 \\ \cline{2-18} 
 & Pl\_loki\_3/8 & 18.49 & 28.35 & \textbf{41.68} & 35.84 & 23.76 & \textbf{14.01} & \textbf{30.13} & 22.52 & 5.02 & 66 & 75.43 & 26.5 & 56.7 & 34.19 & 0.62 & 0.68 \\
 & Salca\_3/8 & \textbf{20.69} & \textbf{28.82} & 40.03 & \textbf{36.59} & \textbf{24.39} & 13.90 & 30.02 & \textbf{22.85} & \textbf{6.34} & \textbf{67} & \textbf{77.64} & \textbf{29.75} & \textbf{56.82} & \textbf{34.99} & 0.72 & 0.79 \\ \cline{2-18} 
 & Pl\_loki\_1/2 & \textbf{19.92} & 28.18 & \textbf{41.26} & 34.91 & \textbf{24.29} & \textbf{14.24} & 29.36 & \textbf{22.82} & 5.85 & 66 & 75.6 & 27 & \textbf{57.53} & 34.38 & 0.77 & 0.93 \\
 & Salca\_1/2 & 19.77 & \textbf{29.12} & 39.76 & \textbf{35.99} & 24.14 & 13.67 & \textbf{30.06} & 22.68 & \textbf{6.25} & \textbf{66.5} & \textbf{76.38} & \textbf{30.5} & 57.4 & \textbf{34.79} & 0.79 & 0.92 \\ \hline
\multirow{6}{*}{Vicuna-v1.5-7B-16k} & Pl\_loki\_1/4 & 13.47 & 26.28 & 34.6 & 22.63 & \textbf{22.15} & 5.82 & 22.71 & 20.74 & 7.61 & 69.5 & 69.15 & 3.5 & 38.03 & 27.4 & 0.62 & 0.74 \\
 & Salca\_1/4 & \textbf{17.22} & \textbf{26.83} & \textbf{35.49} & \textbf{24.28} & 21.5 & \textbf{8.26} & \textbf{27.09} & \textbf{22.46} & \textbf{11.14} & \textbf{69.5} & \textbf{72.2} & \textbf{5.5} & \textbf{39.85} & \textbf{29.33} & 0.75 & 0.82 \\ \cline{2-18} 
 & Pl\_loki\_3/8 & \textbf{18.84} & 26.25 & \textbf{35.45} & 22.99 & 21.06 & 7.62 & \textbf{27.39} & 22.6 & 12.55 & 66.18 & \textbf{75.19} & 4.5 & \textbf{40.41} & \textbf{29.31} & 0.65 & 0.77 \\
 & Salca\_3/8 & 18.47 & \textbf{26.26} & 34.77 & \textbf{24.43} & \textbf{21.95} & \textbf{7.64} & 27.17 & \textbf{23.24} & \textbf{12.68} & \textbf{69} & 67.39 & 4.5 & 39.77 & 29.25 & 0.84 & 0.85 \\ \cline{2-18} 
 & Pl\_loki\_1/2 & \textbf{18.15} & 26.02 & 35.01 & 23.19 & 21.25 & 7.71 & \textbf{27.75} & 22.56 & 12.84 & 69 & \textbf{68.57} & 4.5 & \textbf{40.39} & 29 & 0.81 & 0.94 \\
 & Salca\_1/2 & 18.01 & \textbf{26.29} & \textbf{35.17} & \textbf{24.42} & \textbf{21.26} & \textbf{7.74} & 27.62 & \textbf{23.18} & \textbf{13.6} & 69 & 66.85 & 4.5 & 39.88 & \textbf{29.04} & 0.80 & 0.90 \\ \hline
\multirow{6}{*}{ChatGLM3-6B-32k} & loki\_1/4 & 23.14 & 41.84 & \textbf{49.56} & \textbf{57.66} & 42.47 & \textbf{37.06} & 34.22 & 23.48 & \textbf{17.66} & \textbf{78.5} & \textbf{84.23} & 95.5 & 52.23 & 48.46 & 0.49 & 0.67 \\
 & Salca\_1/4 & \textbf{27.68} & \textbf{42.21} & 49.2 & 55.23 & \textbf{44.57} & 36.05 & \textbf{34.95} & \textbf{24.03} & 15.47 & 67.5 & 83.43 & \textbf{97.5} & \textbf{52.45} & \textbf{48.48} & 0.61 & 0.76 \\ \cline{2-18} 
 & loki\_3/8 & 24.56 & \textbf{43.74} & 48.84 & \textbf{55.31} & 41.96 & 35.63 & \textbf{35.2} & 24.26 & \textbf{15.94} & \textbf{79} & \textbf{83.7} & 99.5 & \textbf{53.92} & 49.15 & 0.69 & 0.79 \\
 & Salca\_3/8 & \textbf{27.21} & 42.72 & \textbf{50.63} & 54.23 & \textbf{46.54} & \textbf{38.08} & 35.02 & \textbf{24.53} & 15.71 & 77.5 & 83.33 & \textbf{98} & 53.08 & \textbf{49.74} & 0.74 & 0.82 \\ \cline{2-18} 
 & loki\_1/2 & 25.73 & \textbf{42.88} & 50.97 & 54.79 & 44.08 & 38.07 & \textbf{36.86} & 24.14 & 15.87 & 79 & 83.53 & \textbf{99.5} & 53.04 & 49.88 & 0.82 & 0.91 \\
 & Salca\_1/2 & \textbf{26.16} & 42.52 & \textbf{51.32} & \textbf{54.93} & \textbf{45.9} & \textbf{39.87} & 36.09 & \textbf{24.33} & \textbf{16.27} & 79 & 83.53 & 99 & \textbf{52.63} & \textbf{50.12} & 0.80 & 0.92 \\ \hline
\end{tabular}%
}
\end{table*}

\subsection{Architecture Evaluation}
\noindent\textbf{Methodology:} For hardware design, we target $s_f = 1/2$, which is compatible with other feature sparsity, and organize on-chip memory for a maximum context length of 64K. A single HBM2 serves as off-chip DRAM, providing a peak bandwidth of 512 GB/s. Following performance model, we set computing parallelism to $16$ for pre-computing and $1$ for attention. We implemented the RTL and utilized VCS to simulate cycle counts for each application. To evaluate area and power, we synthesized the design using Synopsys Design Compiler with TSMC 28nm standard cell library. HBM runs at 1 GHz. Computing core meets sub-1ns timing, but we cap it at 500 MHz—higher frequency would lead to data starvation. For HBM power evaluation, we implemented same architecture on an Alveo U280 Data Center Accelerator Card\cite{U280} and measured HBM power under identical access frequency and data patterns.

We compare Salca with A100 GPU. Elapsed time is measured by inserting \texttt{torch.cuda.synchronize} at beginning and end of attention decoding. For power evaluation, we use \texttt{pynvml} library. Our baseline configurations include GPU\_D and ASIC\_D, which serve as dense computing without sparsity, and GPU\_S, which adopts the proposed compression method.

\begin{figure}[t]
\centering
\begin{subfigure}[b]{0.48\columnwidth}
\centering
\includegraphics[width=\linewidth]{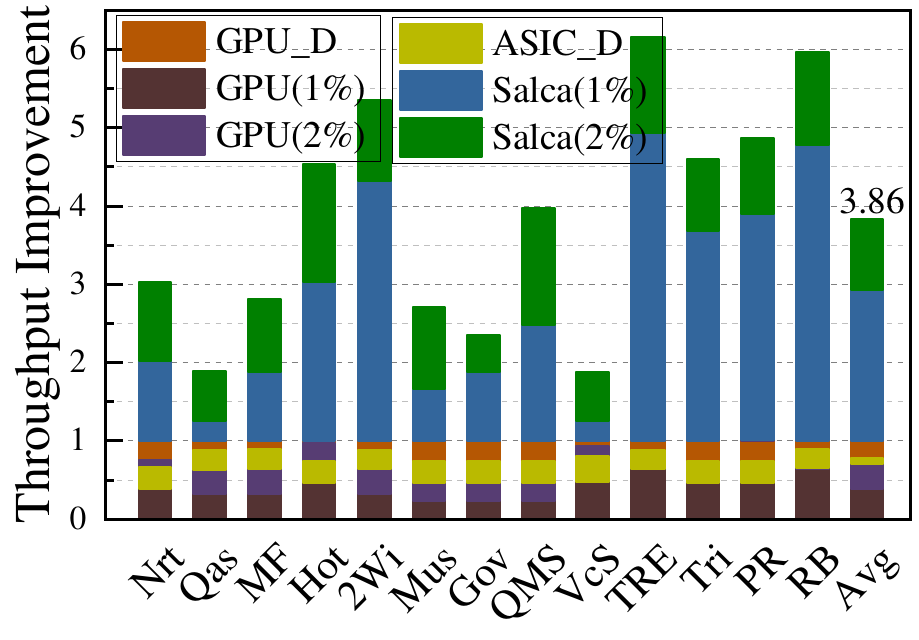}
\caption{Throughput Improvement}
\label{fig:Throughput}
\end{subfigure}
\hfill
\begin{subfigure}[b]{0.48\columnwidth}
\centering
\includegraphics[width=\linewidth]{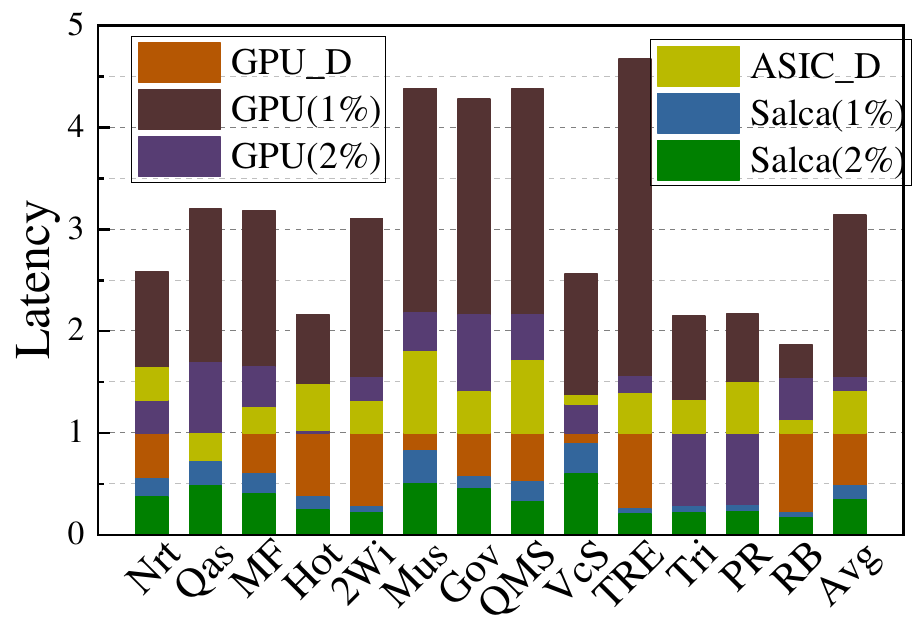}
\caption{Latency}
\label{fig:latency}
\end{subfigure}
\caption{Throughput and Latency.}
\label{fig:main}
\end{figure}

\noindent\textbf{Throughput:} We evaluate throughput of Salca and GPU during decoding under dense computing and diverse accuracy loss(1\% and 2\%). Fig.\ref{fig:Throughput} and Fig.\ref{fig:latency} reports normalized throughput and latency relative to GPU\_D. During decoding, GPU\_D is bottlenecked by compute-memory imbalance, achieving an average throughput of only 1.14 TFLOPS. Introducing sparsity leads to a counterintuitive decrease in GPU throughput. Further analysis shows that loading just $10\%$ of selected K/V vectors accounts for $90\%$ of total execution time. This stems from ineffective caching and prefetching, along with lower HBM read efficiency. As sparsity increases, frequency of discrete accesses diminishes, and throughput gradually improves. This exposes performance limitations of GPUs when processing irregular sparse attention. Constrained by bandwidth, baseline ASIC\_D achieves a throughput of 0.9 TFLOPS, slightly lower than that of GPU\_D. This is because A100 employs five HBM2E devices for 2 TB/s bandwidth, which is $4\times$ that of our single HBM2. However, through dedicated pipeline and memory access optimizations, ASIC can achieve significant advantages. Specifically, Salca($1\%$) and Salca($2\%$) achieve performance gains of $2.81\times$ and $3.82\times$ over GPU\_D. Compared to GPU with same loss, Salca delivers speedups of $7.1\times$ and $5.3\times$, respectively. This validates that Salca has exceptional efficiency for sparse attention under bandwidth-constrained conditions.

\begin{figure}[!htbp]
\centering
\begin{subfigure}[b]{0.48\columnwidth}
\centering
\includegraphics[width=\linewidth]{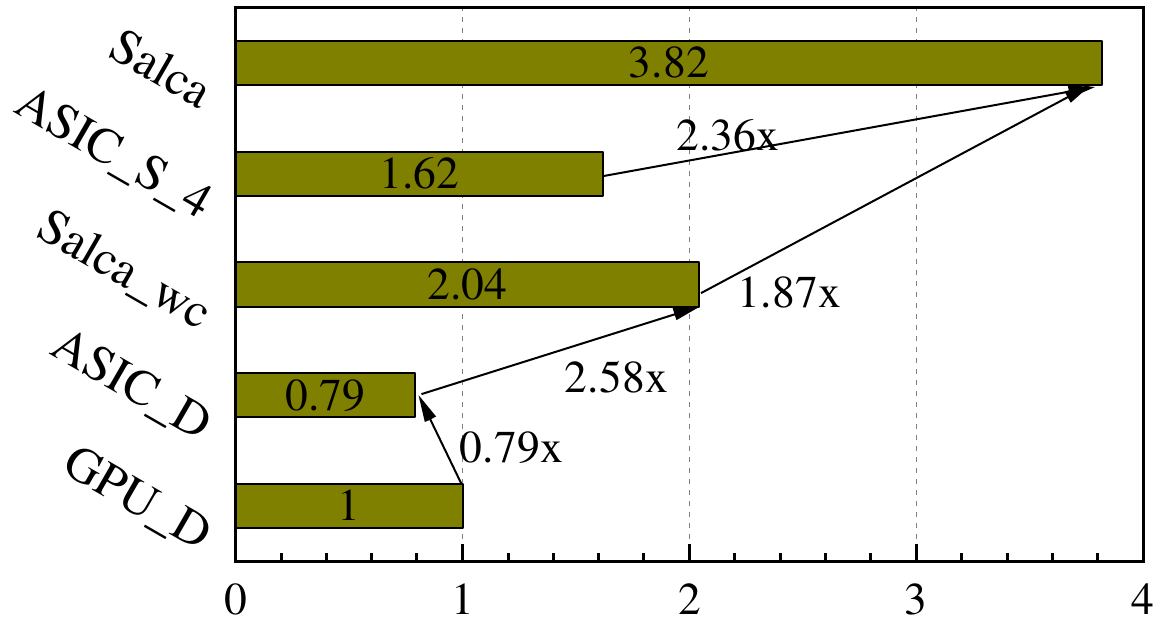}
\caption{Throughput Gain breakdown}
\label{fig:Throughput_Gain_breakdown}
\end{subfigure}
\hfill
\begin{subfigure}[b]{0.48\columnwidth}
\centering
\includegraphics[width=\linewidth]{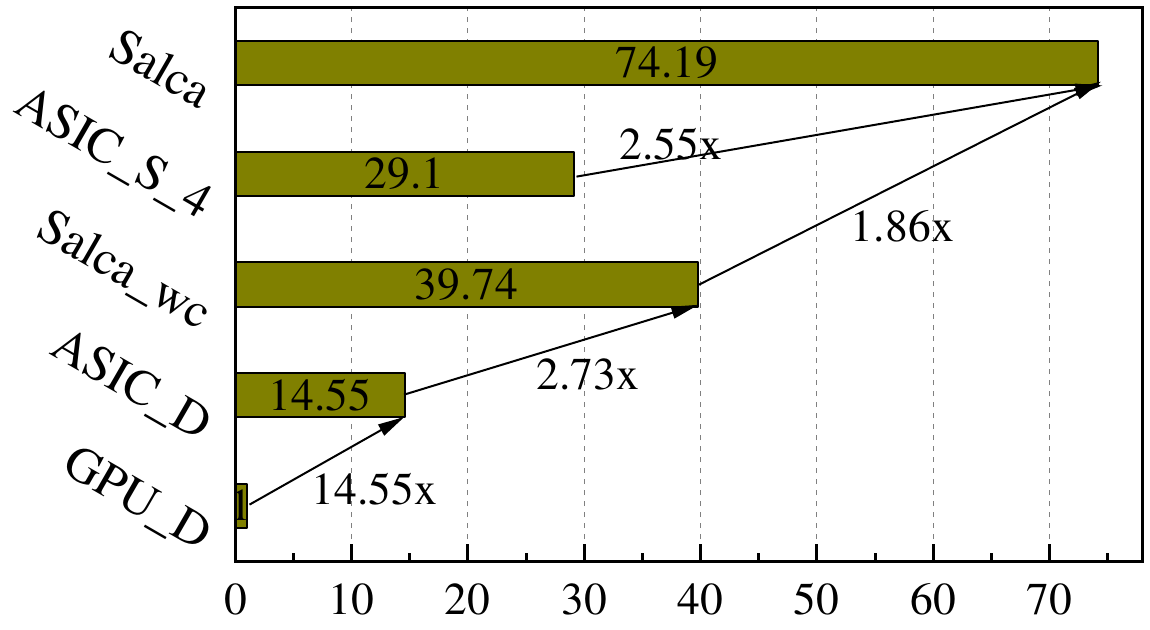}
\caption{Energy Gain breakdown}
\label{fig:Energy_Gain_breakdow}
\end{subfigure}
\caption{Gain breakdown.}
\label{fig:main}
\end{figure}

Fig.\ref{fig:Throughput_Gain_breakdown} illustrates breakdown of throughput gains, which primarily stem from two aspects. \textcircled{1} \textbf{Sparse method gain:} Our sparse method delivers $2.58\times$ speedup over ASIC\_D due to reduced computation. \textcircled{2} \textbf{Architecture optimization gain:} By incorporating conflict elimination mechanism with range of $128$, we reduced HBM channel conflict rate from $2.18$ to $1.17$. This delivers an additional $1.87\times$ performance boost. We also compare with conventional 4-bit quantization schemes (ASIC\_S\_4), as in Energon and Sanger. Their pre-computing loads $4\times$ more data than Salca. Increased bandwidth demand limits the minimum retention rate to $13\%$, making it unsuitable for high sparsity scenarios and resulting in only a modest $1.62\times$ speedup. This reveals data supply bottleneck of conventional schemes when processing high sparsity. In contrast, Salca overcomes this limitation by applying aggressive dual compression. It supports sparsity as low as $5.8\%$ with lower bandwidth consumption and delivers more pronounced improvement.

\begin{figure}[t]
\centering
\begin{subfigure}[b]{0.48\columnwidth}
\centering
\includegraphics[width=\linewidth]{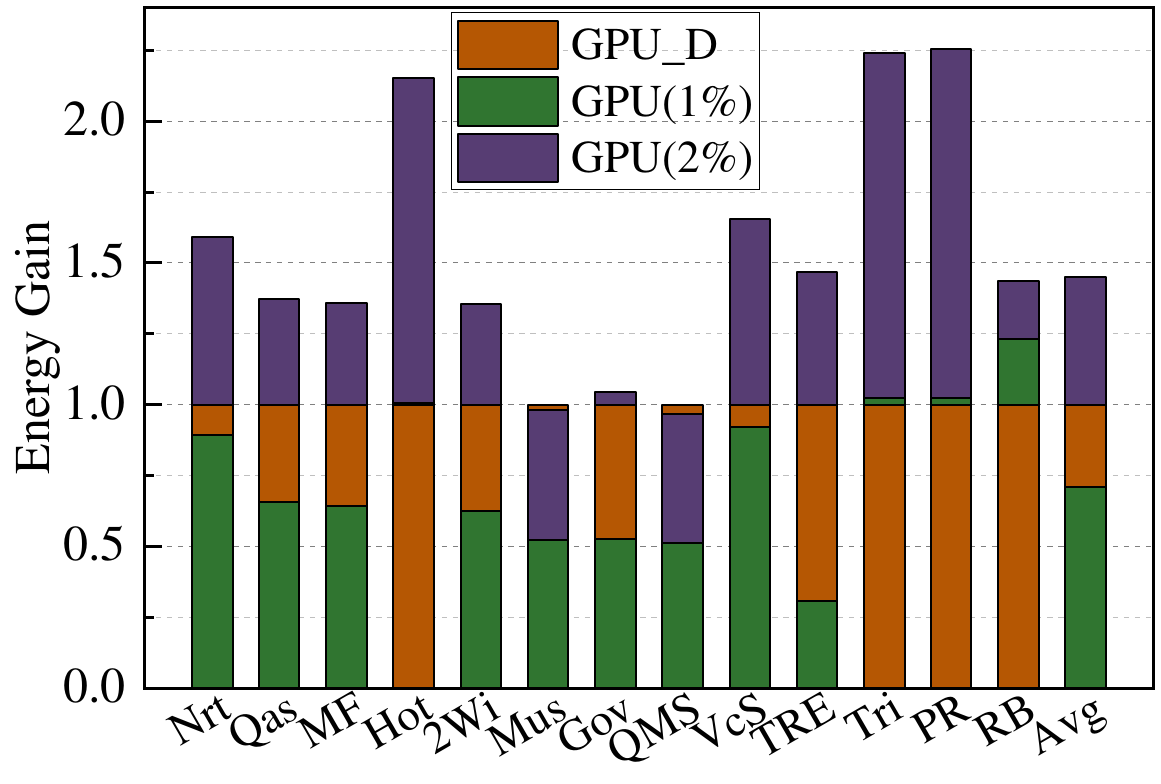}
\caption{Energy Gain of GPU}
\label{fig:Energy_Gain_GPU}
\end{subfigure}
\hfill
\begin{subfigure}[b]{0.48\columnwidth}
\centering
\includegraphics[width=\linewidth]{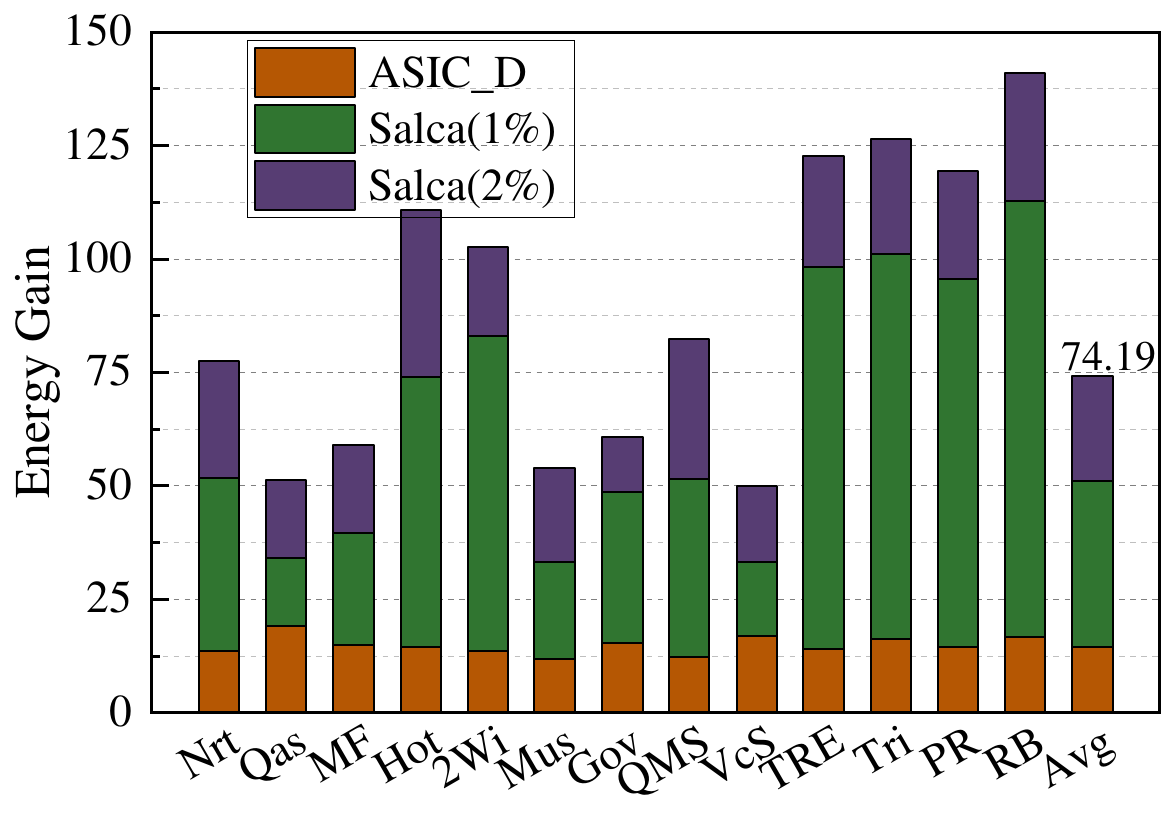}
\caption{Energy Gain of ASIC}
\label{fig:Energy_Gain_asic}
\end{subfigure}
\caption{Energy Gain.}
\label{fig:main}
\end{figure}

\noindent\textbf{Area, Power and Energy:} Tab.\ref{tab:area_and_power} details power and area breakdown of Salca. It occupies a total area of $6.4$ mm² and consumes $0.933$ W. Benefiting from effective compression, pre-computing module incurs minimal overhead, accounting for only $5\%$ of total area and $13\%$ of total power. Primary overhead stems from on-chip buffer, which contributes $87\%$ area and $68\%$ power. HBM power is $9.85$ W, bringing device consumption to $10.783$ W. Energy efficiency results normalized to GPU\_D are presented in Fig.~\ref{fig:Energy_Gain_GPU} and Fig.~\ref{fig:Energy_Gain_asic} (separate plots due to scale disparity). GPU efficiency degrades at low sparsity due to prolonged latency. Despite no throughput gain at higher sparsity, reduced computation and data movement lower dynamic power, resulting in a $1.45\times$ average efficiency improvement. ASIC intrinsically excels in energy efficiency. ASIC\_D delivers $14.55\times$ gains over GPU\_D. Salca($1\%$) and Salca($2\%$) achieve $51.11\times$ and $74.19\times$ energy improvements with co-optimized algorithm and architecture. Fig.\ref{fig:Energy_Gain_breakdow} decomposes these gains: algorithm and conflict elimination independently contribute $2.73\times$ and $1.86\times$ improvements. Furthermore, Salca outperforms ASIC\_S\_4 by $2.55\times$, demonstrating superior efficiency in processing sparse attention.

\begin{table}[t]
\centering
\setlength{\abovecaptionskip}{0pt}
\caption{Area and Power}
\label{tab:area_and_power}
\resizebox{\columnwidth}{!}{%
\begin{tabular}{cccc}
\hline
 & Stage & Area(mm$^2$) & Power(W) \\ \hline
\multirow{3}{*}{Pre-Computing} & Relevance Estimation & 0.0734 & 0.0275 \\
 & Threshold Locating & 0.1973 & 0.088 \\
 & Traverse & 0.04485 & 0.007 \\ \hline
\multirow{2}{*}{Attention} & QKmul & 0.2077 & 0.0659 \\
 & SM\_PVmul & 0.319 & 0.1116 \\ \hline
Buffer & - & 5.56 & 0.633 \\ \hline
Total & - & 6.4 & 0.933 \\ \hline
\end{tabular}%
}
\end{table}

\begin{table*}[t]
\centering
\setlength{\abovecaptionskip}{0pt}
\caption{Comparison results with existing accelerators}
\label{tab:Comp_result_with_existing}
\resizebox{\textwidth}{!}{%
\begin{tabular}{cccccccccccccc}
\hline
 & \multirow{2}{*}{\begin{tabular}[c]{@{}c@{}}Evaluated\\ Maxlen\end{tabular}} & \multirow{2}{*}{\begin{tabular}[c]{@{}c@{}}Mem\\ Saving\end{tabular}} & \multirow{2}{*}{\begin{tabular}[c]{@{}c@{}}Comp\\ Saving\end{tabular}} & \multirow{2}{*}{\begin{tabular}[c]{@{}c@{}}Top-K \\ Comp\end{tabular}} & \multirow{2}{*}{\begin{tabular}[c]{@{}c@{}}Tech\\ {[}nm{]}\end{tabular}} & \multirow{2}{*}{\begin{tabular}[c]{@{}c@{}}Freq\\ {[}Hz{]}\end{tabular}} & \multirow{2}{*}{\begin{tabular}[c]{@{}c@{}}Area\\ {[}mm$^2${]}$^{\text{1}}$\end{tabular}} & \multicolumn{2}{c}{Power{[}W{]}} & \multirow{2}{*}{\begin{tabular}[c]{@{}c@{}}Throup\\ {[}GOPS{]}\end{tabular}} & \multicolumn{2}{c}{Energy Effi {[}GOPS/W{]$^{\text{3}}$}} & \multirow{2}{*}{\begin{tabular}[c]{@{}c@{}}Area Effi\\ {[}GOPS/mm$^2${]}$^{\text{3}}$\end{tabular}} \\ \cline{9-10} \cline{12-13}
 &  &  &  &  &  &  &  & Core & IO$^{\text{2}}$ &  & Core & IO &  \\ \hline
A3\cite{A3} & 320 & 40\% & 40\% & O(n) & 40 & 1G & 2.08/3.50 & 0.205 & 0.617/9.83 & 221 & 1863 & 300/39 & 217/63 \\
ELSA\cite{ELSA} & 512 & 84\% & 73\% & O(nlogn) & 40 & 1G & 1.26/3.12 & 0.969 & 0.525/9.8 & 1090 & 1944 & 1004/193 & 1756/349 \\
Sanger\cite{Sanger} & 4K & 65\% & 76\% & O(n) & 55 & 500M & 16.9/4.38$^{\text{4}}$ & 2.76 & -/1.22 & 2285/36 & 2342/37 & -/52 & 522/8 \\
DOTA\cite{DOTA} & 4K & 85\% & 80\% & O(nlogn) & 22 & 1G & 4.44/9.69 & 3.02 & -/39.32 & 4905/1226 & 817/204 & -/14 & 683/127 \\
Energon\cite{Energon} & 1K & 71\% & 77\% & O(n) & 45 & 1G & 4.20/4.12 & 0.32 & 2.4/9.83 & 1153 & 7007/3504$^{\text{5}}$ & 450/113 & 709/280 \\
SpAtten\cite{Spatten} & 1K & 60\% & 67\% & O(nlogn) & 40 & 1G & 1.55/3.26 & 0.325 & 0.617/9.83 & 360 & 1915 & 447/62 & 474/110 \\
FACT\cite{FACT} & 512 & - & 79\% & O(nlogn) & 28 & 500M & 6.03/8.53 & 0.337 & -/19.66 & 928 & 2754 & -/46 & 154/108 \\
SOFA\cite{SOFA} & 4K & 79\% & 82\% & O(nlogn) & 28 & 1G & 5.69/8.19 & 0.95 & 2.45/19.66 & 24428/191 & 25708/201 & 7183/9 & 4292/25 \\
Salca & 64K & 87\% & 87\% & O(n) & 28 & 500M & 6.4 & 0.933 & 9.83 & 4350 & 4662 & 403 & 680 \\ \hline
\end{tabular}%
}
\begin{flushleft}
\footnotesize
$^{\text{1}}$ Area in LCS are scaled to 28nm; $^{\text{2}}$ IO Power in LCS are scaled to 28nm; $^{\text{3}}$ Scaled to 28nm; $^{\text{4}}$ Use hyperparameter as threshold, do not need buffer; $^{\text{5}}$ ODF strategy breakdown incurs a $1.3\times$ efficiency penalty and data accesses required for the second filtering stage become discrete short-burst HBM reads, resulting in at least $35\%$ read efficiency loss
\end{flushleft}
\end{table*}

\begin{table*}[]
\centering
\setlength{\abovecaptionskip}{0pt}
\caption{Accuracy under different quantization method}
\label{tab:mini_quant_width_test}
\resizebox{\textwidth}{!}{%
\begin{tabular}{ccccccccccccccccccccccccccc}
\hline
\multicolumn{1}{l}{} & \multicolumn{13}{c|}{LongChat-v1.5-7B-32k} & \multicolumn{13}{c}{Vicuna-v1.5-7B-16k} \\ \cline{2-27} 
{\color[HTML]{000000} } & {\color[HTML]{000000} Nrt} & {\color[HTML]{000000} Qas} & {\color[HTML]{000000} MF} & {\color[HTML]{000000} Hot} & {\color[HTML]{000000} 2Wi} & {\color[HTML]{000000} Mus} & {\color[HTML]{000000} Gov} & {\color[HTML]{000000} QMS} & {\color[HTML]{000000} VcS} & {\color[HTML]{000000} TRE} & {\color[HTML]{000000} Tri} & {\color[HTML]{000000} PR} & {\color[HTML]{000000} RB} & Nrt & Qas & MF & Hot & 2Wi & Mus & Gov & QMS & VcS & TRE & Tri & PR & RB \\ \hline
{\color[HTML]{000000} baseline} & {\color[HTML]{000000} 19.51} & {\color[HTML]{000000} 27.51} & {\color[HTML]{000000} 31.86} & {\color[HTML]{000000} 29.46} & {\color[HTML]{000000} 22.2} & {\color[HTML]{000000} 11.58} & {\color[HTML]{000000} 25.43} & {\color[HTML]{000000} 21.82} & {\color[HTML]{000000} 2.5} & {\color[HTML]{000000} 65.5} & {\color[HTML]{000000} 56.24} & {\color[HTML]{000000} 19.75} & {\color[HTML]{000000} 44.37} & 13.73 & 21.68 & 25.91 & 17.74 & 16.07 & 7.57 & 21.27 & 19.1 & 5.51 & 66.5 & 56.93 & 4 & 34.45 \\
{\color[HTML]{000000} k\_1} & {\color[HTML]{000000} 11.53} & {\color[HTML]{000000} 18.45} & {\color[HTML]{000000} 19.13} & {\color[HTML]{000000} 19.92} & {\color[HTML]{000000} 12.11} & {\color[HTML]{000000} 5.84} & {\color[HTML]{000000} 6.1} & {\color[HTML]{000000} 10.97} & {\color[HTML]{000000} 5.66} & {\color[HTML]{000000} 47.5} & {\color[HTML]{000000} 41.49} & {\color[HTML]{000000} 4.67} & {\color[HTML]{000000} 21.94} & 10.6 & 11.69 & 9.95 & 12.35 & 12.74 & 4.18 & 2.57 & 9.31 & 8.48 & 43.5 & 25.96 & 1 & 23.97 \\
{\color[HTML]{000000} \textbf{k\_2\_asy}} & {\color[HTML]{000000} 19.26} & {\color[HTML]{000000} 27.28} & {\color[HTML]{000000} 34.61} & {\color[HTML]{000000} 28.89} & {\color[HTML]{000000} 23.07} & {\color[HTML]{000000} 11.58} & {\color[HTML]{000000} 27.3} & {\color[HTML]{000000} 22.44} & {\color[HTML]{000000} 2.78} & {\color[HTML]{000000} 66} & {\color[HTML]{000000} 60.18} & {\color[HTML]{000000} 25} & {\color[HTML]{000000} 50.91} & 13.54 & 21.63 & 25.74 & 19.41 & 16.04 & 8.39 & 23.01 & 18.91 & 6.19 & 67.5 & 57.07 & 4 & 34.98 \\
{\color[HTML]{000000} k\_2\_sym} & {\color[HTML]{000000} 14.05} & {\color[HTML]{000000} 13.87} & {\color[HTML]{000000} 20.92} & {\color[HTML]{000000} 27.07} & {\color[HTML]{000000} 17.91} & {\color[HTML]{000000} 8.89} & {\color[HTML]{000000} 8.98} & {\color[HTML]{000000} 19.69} & {\color[HTML]{000000} 2.03} & {\color[HTML]{000000} 53.5} & {\color[HTML]{000000} 43.49} & {\color[HTML]{000000} 7.97} & {\color[HTML]{000000} 15.73} & 9.98 & 16.24 & 18.49 & 14.37 & 16.33 & 4.76 & 11.72 & 17.28 & 4.13 & 63.25 & 43.15 & 4.5 & 23.99 \\
{\color[HTML]{000000} k\_3\_asy} & {\color[HTML]{000000} 18.94} & {\color[HTML]{000000} 26.01} & {\color[HTML]{000000} 32.73} & {\color[HTML]{000000} 28.8} & {\color[HTML]{000000} 22.2} & {\color[HTML]{000000} 11.5} & {\color[HTML]{000000} 26.6} & {\color[HTML]{000000} 22.24} & {\color[HTML]{000000} 2.61} & {\color[HTML]{000000} 65.5} & {\color[HTML]{000000} 56.37} & {\color[HTML]{000000} 22.5} & {\color[HTML]{000000} 45.63} & 14.03 & 21.8 & 23.93 & 17.52 & 15.64 & 7.81 & 21.85 & 19.52 & 4.75 & 5.75 & 57.48 & 4.5 & 37.19 \\
{\color[HTML]{000000} k\_3\_sym} & {\color[HTML]{000000} 19.51} & {\color[HTML]{000000} 28.11} & {\color[HTML]{000000} 32.12} & {\color[HTML]{000000} 29.38} & {\color[HTML]{000000} 23.04} & {\color[HTML]{000000} 11.66} & {\color[HTML]{000000} 26.46} & {\color[HTML]{000000} 22.35} & {\color[HTML]{000000} 2.4} & {\color[HTML]{000000} 65.5} & {\color[HTML]{000000} 57.57} & {\color[HTML]{000000} 22.5} & {\color[HTML]{000000} 46.35} & 13.79 & 21.9 & 25.52 & 18.26 & 15.46 & 8.51 & 21.66 & 18.93 & 5.82 & 65.5 & 57.26 & 4 & 36.81 \\
{\color[HTML]{000000} k\_msb2\_asy} & {\color[HTML]{000000} 18.42} & {\color[HTML]{000000} 26.51} & {\color[HTML]{000000} 35.59} & {\color[HTML]{000000} 29.75} & {\color[HTML]{000000} 22.84} & {\color[HTML]{000000} 9.87} & {\color[HTML]{000000} 24.68} & {\color[HTML]{000000} 20.28} & {\color[HTML]{000000} 2.75} & {\color[HTML]{000000} 64} & {\color[HTML]{000000} 54.28} & {\color[HTML]{000000} 23.5} & {\color[HTML]{000000} 23.5} & 14.2 & 19.26 & 23.47 & 17.18 & 16.78 & 8.52 & 19.99 & 19.26 & 5.77 & 67 & 53.45 & 4 & 34.32 \\
{\color[HTML]{000000} k\_msb2\_sym} & {\color[HTML]{000000} 17.6} & {\color[HTML]{000000} 22.3} & {\color[HTML]{000000} 26.39} & {\color[HTML]{000000} 28.57} & {\color[HTML]{000000} 20.28} & {\color[HTML]{000000} 9.83} & {\color[HTML]{000000} 17.75} & {\color[HTML]{000000} 21.13} & {\color[HTML]{000000} 2.55} & {\color[HTML]{000000} 62.5} & {\color[HTML]{000000} 52.4} & {\color[HTML]{000000} 23.42} & {\color[HTML]{000000} 34.52} & 12.45 & 18.67 & 19.93 & 17.57 & 17.12 & 7.23 & 15.47 & 18.65 & 6.51 & 67.5 & 51.58 & 4.25 & 35.52 \\
{\color[HTML]{000000} k\_msb3\_asy} & {\color[HTML]{000000} 18.62} & {\color[HTML]{000000} 27.62} & {\color[HTML]{000000} 31.68} & {\color[HTML]{000000} 29.8} & {\color[HTML]{000000} 22.07} & {\color[HTML]{000000} 11.06} & {\color[HTML]{000000} 25.55} & {\color[HTML]{000000} 21.7} & {\color[HTML]{000000} 2.78} & {\color[HTML]{000000} 65} & {\color[HTML]{000000} 57.3} & {\color[HTML]{000000} 20.75} & {\color[HTML]{000000} 43.52} & 14.61 & 20.84 & 24.1 & 18.52 & 16.97 & 7.29 & 20.75 & 19.5 & 5.71 & 66 & 55.66 & 4 & 35.7 \\
{\color[HTML]{000000} k\_msb3\_sym} & {\color[HTML]{000000} 20.25} & {\color[HTML]{000000} 25.66} & {\color[HTML]{000000} 30.71} & {\color[HTML]{000000} 28.27} & {\color[HTML]{000000} 22.23} & {\color[HTML]{000000} 10.72} & {\color[HTML]{000000} 24.41} & {\color[HTML]{000000} 22.04} & {\color[HTML]{000000} 2.51} & {\color[HTML]{000000} 65} & {\color[HTML]{000000} 57.68} & {\color[HTML]{000000} 19.17} & {\color[HTML]{000000} 41} & 13.6 & 20.31 & 23.71 & 17.59 & 16.88 & 6.85 & 20.83 & 18.87 & 6.13 & 65.5 & 55.89 & 4.5 & 34.89 \\
{\color[HTML]{000000} q\_1\_sym} & {\color[HTML]{000000} 3.85} & {\color[HTML]{000000} 6.23} & {\color[HTML]{000000} 7.69} & {\color[HTML]{000000} 11.51} & {\color[HTML]{000000} 6.3} & {\color[HTML]{000000} 1.73} & {\color[HTML]{000000} 1.07} & {\color[HTML]{000000} 4.17} & {\color[HTML]{000000} 2.72} & {\color[HTML]{000000} 29.25} & {\color[HTML]{000000} 5.73} & {\color[HTML]{000000} 0} & {\color[HTML]{000000} 10.91} & 3.22 & 6.7 & 4.21 & 5.79 & 5.87 & 0.86 & 0.87 & 2.87 & 2.31 & 19.5 & 4.32 & 2 & 10.49 \\
{\color[HTML]{000000} q\_2\_sym} & {\color[HTML]{000000} 18.34} & {\color[HTML]{000000} 23.14} & {\color[HTML]{000000} 29.97} & {\color[HTML]{000000} 29.54} & {\color[HTML]{000000} 21.15} & {\color[HTML]{000000} 11.04} & {\color[HTML]{000000} 19.93} & {\color[HTML]{000000} 20.02} & {\color[HTML]{000000} 2.42} & {\color[HTML]{000000} 59.75} & {\color[HTML]{000000} 56.52} & {\color[HTML]{000000} 19.62} & {\color[HTML]{000000} 34.24} & 11.27 & 16.08 & 18.69 & 16.76 & 16.72 & 4.51 & 13.54 & 15.35 & 5.89 & 61 & 51.12 & 4.46 & 24.87 \\
{\color[HTML]{000000} \textbf{q\_3\_sym}} & {\color[HTML]{000000} 18.83} & {\color[HTML]{000000} 28.1} & {\color[HTML]{000000} 34.55} & {\color[HTML]{000000} 30.39} & {\color[HTML]{000000} 23.6} & {\color[HTML]{000000} 11.2} & {\color[HTML]{000000} 27.76} & {\color[HTML]{000000} 21.9} & {\color[HTML]{000000} 2.93} & {\color[HTML]{000000} 66} & {\color[HTML]{000000} 59.15} & {\color[HTML]{000000} 24.75} & {\color[HTML]{000000} 48.05} & 13.77 & 23.03 & 28.63 & 20.09 & 17.73 & 7.85 & 23.84 & 18.77 & 6.5 & 68 & 59.21 & 4 & 35.28 \\
{\color[HTML]{000000} q\_4\_sym} & {\color[HTML]{000000} 19.32} & {\color[HTML]{000000} 28.49} & {\color[HTML]{000000} 33.9} & {\color[HTML]{000000} 29.79} & {\color[HTML]{000000} 23.25} & {\color[HTML]{000000} 11.25} & {\color[HTML]{000000} 26.88} & {\color[HTML]{000000} 22.37} & {\color[HTML]{000000} 2.82} & {\color[HTML]{000000} 66} & {\color[HTML]{000000} 58.85} & {\color[HTML]{000000} 25.92} & {\color[HTML]{000000} 50.23} & 13.88 & 21.71 & 25.93 & 18.85 & \textit{17.22} & 7.42 & 23.59 & 19.11 & 5.98 & 67.5 & 57.97 & 4 & 34.86 \\ \hline
\end{tabular}%
}
\end{table*}

\subsection{Comparison with existing Accelerators}
Decoding of LCS poses severe challenges to hardware design. First, surging on-chip memory inflate die area. Second, ineffective on-chip reuse triggers frequent HBM accesses, significantly increasing power consumption. Existing accelerators are primarily designed and evaluated in SCS. Their performance metrics cannot truly reflect processing capability in LCS. To ensure a fair comparison, we conducted an equivalent evaluation under long context workloads. \textbf{Throughput:} Some accelerators optimize prefilling by exploiting multi-query parallelism. However, decoding processes one query at a time, leading to resource underutilization. Therefore, $\operatorname{T}_{LCS} = \frac{\operatorname{T}_{SCS}}{\mathrm{Parallelism}_q}$. Directly evaluating the impact of data supply on throughput in LCS is infeasible. Instead, we assume sufficient data supply and evaluate this factor indirectly through energy efficiency. \textbf{Power:} We take multiplier count of Salca attention $M_{Salca}$ and IO power $P_{SalcaIO}$ as baseline. It reflects IO power when $M_{Salca}$ multipliers continuously fetch data from HBM at 500 MHz. If multiplier count or frequency increases, IO bandwidth and power will also change linearly. Therefore, $P_{HBM} = \frac{freq}{500} \cdot \frac{Mult}{M_{Salca}} \cdot P_{SalcaIO}$. \textbf{Area:} To support a context length of 64K, all accelerators must incorporate at least one buffer with a capacity of 128K data into pipeline. Therefore, $A_{LCS}=A_{SCS}+A_{buf}$.

Based on this transformation, we analyze performance of existing accelerators, as shown in Tab.\ref{tab:Comp_result_with_existing}. We mark the performance changes in LCS after slash. Salca achieves at least 3.5$\times$ throughput improvement over prior accelerators. SOFA, FACT, Sanger, and DOTA are designed for prefilling. They suffer significant performance drops. Our superior performance is mainly attributed to three aspects. \textcircled{1} Higher sparsity in LCS: Tests on ChatGLM3 show that average retention rate can reach 6.22\% with negligible accuracy loss, providing greater acceleration potential. Meanwhile, Salca can support high sparsity demands. \textcircled{2} Dual compression reduces pre-computing overhead: As shown in columns 3 and 4 of Tab.\ref{tab:Comp_result_with_existing}, it achieves excellent compression in both memory access and computing. Saved memory bandwidth can also be released to support stronger computing capabilities. \textcircled{3} Top-K optimization eliminates hardware bottlenecks. It reduces decoding complexity to $O(n)$. Energon computes threshold based on mean, while Sanger sets it via hyperparameters. Both approaches eliminate performance bottleneck of exact Top-K. However, these methods lack precise control over sparsity. Actual performance hinges on accuracy of threshold selection. In contrast, our sorting mechanism circumvents the uncertainty while maintaining equivalent complexity.

Salca also shows significant efficiency gains. For prior accelerators, energy efficiency drops sharply. Frequent off-chip accesses lead to higher IO power. In contrast, Salca achieves at least 1.33$\times$ better energy efficiency at core and 2.08$\times$ at device. Beyond high throughput, Salca's energy advantage stems from superior hardware utilization. Existing accelerators suffer from data starvation, hurting energy efficiency. Salca avoids this by co-designing compute and memory. It uses available bandwidth as design constraint. This ensures that computing and bandwidth are well matched with high hardware and bandwidth utilization. In addition, Salca delivers at least 1.97$\times$ higher area efficiency, mainly due to high throughput. This demonstrates architectural advantages of Salca.

\subsection{Design Space Exploration}
\noindent\textbf{Minimum Quantization Width Exploration:} To determine minimum quantization width required for sparse pattern selection, we evaluate Query and Keys under different quantization strategies and bit-widths. Top $10\%$ selection results under full precision serve as baseline. Exploration consists of two stages. First, we maintain Query in full precision and test nine quantization schemes for Keys, as shown in Tab.\ref{tab:mini_quant_width_test}, including: (1) 1-bit: sign bit only, as in Peng et al.~\cite{DAC_1bit_quant}; (2) 2/3-bit symmetric and asymmetric quantization; (3) MSB\_2/3: quantizes Key to INT8 first, then retains the most significant $2$/$3$ bits, similar to Energon\cite{Energon}. Experimental results indicate that $2$-bit asymmetric quantization yields performance closest to baseline. After quantizing Keys, we further test symmetric quantization for Query from $1$-bit to $4$-bit. We find that $3$-bit achieves best trade-off between accuracy and computing overhead. This scheme also demonstrats excellent performance on ChatGLM3, as evidenced in Tab.~\ref{tab:accuracy_comparison}, validating its generality.

\section{Related Work}
\textbf{Sparse Attention Algorithm:} Various optimizations have been proposed to reduce similarity computing overhead \cite{Lserve} \cite{correia2019adaptively} \cite{Joint_sparse_quant} \cite{zhao2019explicit}. Quest\cite{Quest}, Double Sparsity\cite{double_sparsity}, and Loki\cite{Loki} only select a subset of feature dimensions from Key. MobA\cite{Moba} and NSA\cite{NSA} use grouping methods for block-level relevance estimation. ELSA\cite{ELSA}, DOTA\cite{DOTA}, and PQCache\cite{Pqcache} rely on hashing or clustering to reduce Key size. However, Salca further integrates ultra-low-precision quantization with feature sparsity and demonstrates their orthogonality based on them. This results in an average of just 0.5 bits per feature, dramatically minimizing pre-computing overhead.

\textbf{Sparse Attention Accelerator:} Numerous works on sparse attention acceleration have adopted hardware-software co-design approaches\cite{A3} \cite{AccelTran} \cite{DAC_1bit_quant} \cite{DOTA} \cite{DTATrans} \cite{Edge-llm}. However, existing designs primarily target short context processing. For example, ELSA\cite{ELSA} supports a maximum of 512 tokens, while A3\cite{A3} only handles 320 tokens. In LCS, memory access patterns are fundamentally different, rendering advantages of these algorithms and architectures ineffective. Salca is the first accelerator designed specifically for LCS, breaking input length limitations of existing accelerators.

\textbf{Prefill-Decode Disaggregation:} Prefilling and decoding exhibit distinct computing and memory characteristics, driving academia and industry toward PD-separation architectures\cite{distserve} \cite{spad} \cite{splitwise} \cite{qin2024mooncake} \cite{liao2026dopd}. Prefilling is compute-bound, whereas decoding is memory-bound. However, existing accelerators often fail to distinguish between these two phases. For instance, Energon\cite{Energon} and SpAtten\cite{Spatten} support both processes indiscriminately, ignoring their inherent differences. This inevitably leads to architectural mismatch. Salca adopts this philosophy and focuses on accelerating decoding through ASIC, achieving superior performance.


\bibliographystyle{ACM-Reference-Format}
\bibliography{sample-base}

\end{document}